\documentclass[structabstract]{aa}  
\usepackage{graphicx}
\usepackage{txfonts}
\def\kms{\hbox{\,km\,s$^{-1}$}}       

\def\degr{\hbox{$^\circ$}}

\begin{document}

\title{K2-106, a system containing a metal-rich planet and a planet of
  lower density
\thanks{The results are partly based on observations obtained
    at the European Southern Observatory at Paranal, Chile in program
    098.C-0860(A). This paper includes data gathered with the 6.5 meter
    Magellan Telescopes located at Las Campanas Observatory, Chile.
    The article is also partly based on observations with the TNG, NOT.
    This work has also made use of
    data from the European Space Agency (ESA) mission Gaia
   ({https://www.cosmos.esa.int/gaia}), processed by the Gaia Data
   Processing and Analysis Consortium (DPAC,
  {https://www.cosmos.esa.int/web/gaia/dpac/consortium}).}}
\authorrunning{Guenther et al.}  
\titlerunning{K2-106: a metal-rich low-density planet}
\author{E.W. Guenther\inst{1,7} 
\and O.~Barrag\'{a}n\inst{2} 
\and F.~Dai\inst{3}
\and D.~Gandolfi\inst{2} 
\and T.~Hirano\inst{4} 
\and M.~Fridlund\inst{5,6,7}
\and L.~Fossati\inst{8}
\and A.~Chau\inst{9}
\and R. Helled \inst{9}
\and J.~Korth\inst{10} 
\and J.~Prieto-Arranz\inst{7,11}
\and D.~Nespral \inst{7,11} 
\and G.~Antoniciello\inst{2}
\and H.~Deeg\inst{7,11} 
\and M.~Hjorth\inst{12}
\and S.~Grziwa\inst{10} 
\and S.~Albrecht\inst{12}
\and A.P.~Hatzes \inst{1} 
\and H.~Rauer\inst{13,14}
\and Sz.~Csizmadia\inst{13}
\and A.M.S.~Smith\inst{13}
\and J.~Cabrera\inst{13} 
\and N.~Narita\inst{15,16,17}
\and P.~Arriagada\inst{18}
\and J.~Burt\inst{3}
\and R.P.~Butler\inst{18}
\and W.D.~Cochran\inst{19} 
\and J.D.~Crane\inst{20}
\and Ph.~Eigm\"uller\inst{13} 
\and A.~Erikson\inst{13}
\and J.A.~Johnson\inst{21}
\and A.~Kiilerich\inst{12}
\and D. Kubyshkina\inst{8}
\and E.~Palle\inst{7,11}
\and C.M.~Persson \inst{6}   
\and M.~P\"atzold\inst{10}
\and S.~Sabotta\inst{1} 
\and B.~Sato\inst{4}
\and St.A.~Shectman \inst{20}
\and J.K.~Teske \inst{18,20}
\and I.B.~Thompson \inst{20}
\and V.~Van~Eylen\inst{5}
\and G.~Nowak\inst{7,11} 
\and A.~Vanderburg \inst{21}
\and J.N.~Winn\inst{22}
\and R.A.~Wittenmyer \inst{23}
}
\institute{Th\"uringer Landessternwarte Tautenburg, Sternwarte 5,
  07778 Tautenburg, Germany 
           \and 
Dipartimento di Fisica, Universit\'a di Torino, Via P. Giuria 1,
I-10125, Torino, Italy
          \and 
Department of Physics and Kavli Institute for Astrophysics and Space
Research, Massachusetts Institute of Technology, Cambridge, MA 02139,
USA
           \and 
Department of Earth and Planetary Sciences, Tokyo Institute of
Technology, 2-12-1 Ookayama, Meguro-ku, Tokyo 152-8551, Japan
           \and 
Leiden Observatory, Leiden University, 2333CA Leiden, The
Netherlands
           \and 
Department of Space, Earth and Environment, Chalmers University of Technology,
Onsala Space Observatory, 439 92 Onsala, Sweden
           \and 
Instituto de Astrof\'\i sica de Canarias (IAC), 38205 La Laguna,
Tenerife, Spain
           \and 
Space Research Institute, Austrian Academy of Sciences,
Schmiedlstrasse 6, 8042, Graz, Austria
          \and
Institute for Computational Science, Center for Theoretical
Astrophysics \& Cosmology, University of Zurich, Winterthurerstr. 190,
CH-8057 Zurich, Switzerland
          \and
Rheinisches Institut f\"ur Umweltforschung an der Universit\"at zu
K\"oln, Aachener Strasse 209, 50931 K\"oln, Germany
          \and 
Departamento de Astrof\'\i sica, Universidad de La Laguna (ULL), 38206
La Laguna, Tenerife, Spain
           \and 
Stellar Astrophysics Centre, Department of Physics and Astronomy,
Aarhus University, Ny Munkegade 120, DK-8000 Aarhus C, Denmark
          \and 
Institute of Planetary Research, German Aerospace Center,
Rutherfordstrasse 2, 12489 Berlin, Germany
          \and 
Center for Astronomy and Astrophysics, TU Berlin,
Hardenbergstr. 36, 10623 Berlin, Germany
           \and 
Department of Astronomy, The University of Tokyo, 7-3-1 Hongo,
Bunkyo-ku, Tokyo 113-0033, Japan
           \and 
Astrobiology Center, NINS, 2-21-1 Osawa, Mitaka, Tokyo 181-8588,
Japan
           \and 
National Astronomical Observatory of Japan, NINS, 2-21-1 Osawa,
Mitaka, Tokyo 181-8588, Japan
           \and 
Carnegie Institution of Washington, Department of
Terrestrial Magnetism, 5241 Broad Branch Road, NW, Washington DC,
20015-1305, USA
          \and 
Department of Astronomy and McDonald Observatory, University of Texas
at Austin, 2515 Speedway, Stop C1400, Austin, TX 78712, USA
          \and 
The Observatories of the Carnegie Institution of
Washington, 813 Santa Barbara Street, Pasadena, CA 91101, USA
          \and
Harvard-Smithsonian Center for Astrophysics, 60
Garden Street, Cambridge, MA 02138, USA
          \and 
Princeton University, Department of Astrophysical Sciences, 4 Ivy
Lane, Princeton, NJ 08540 USA
          \and
University of Southern Queensland, Computational
Science and Engineering Research Centre, Toowoomba QLD Australia
\\
              \email{guenther@tls-tautenburg.de}\\
}
\date{Received March 29, 2017; accepted Sep 15, 2017}
  \abstract 
{}
{Planets in the mass range from 2 to 15 $\rm M_{\oplus}$ are very
  diverse. Some of them have low densities, while others are very dense. 
  By measuring the masses and radii, the mean densities, structure, and 
  composition of the planets are constrained. These
  parameters also give us important information about their formation
  and evolution, and about possible processes for atmospheric loss.}
{We determined the masses, radii, and mean densities for the two
  transiting planets orbiting K2-106. The inner planet has an
  ultra-short period of 0.57 days.  The period of the outer planet is
  13.3 days.}
{Although the two planets have similar masses, their densities are very different.
   For K2-106b we derive $\rm M_b=8.36_{-0.94}^{+0.96}$
  $\rm M_{\oplus}$, $\rm R_b=1.52\pm0.16$\,$\rm R_{\oplus}$ ,
  and a high density of $13.1_{-3.6}^{+5.4}$ $\rm g\,cm^{-3}$.  For
  K2-106c, we find $\rm M_c=5.8_{-3.0}^{+3.3}$ $M_{\oplus}$, $\rm
  R_c=2.50_{-0.26}^{+0.27}$\,$\rm R_{\oplus}$ and a relatively low
  density of $2.0_{-1.1}^{+1.6}$ $\rm g\,cm^{-3}$.  }
{Since the system contains two planets of almost the same mass, but
  different distances from the host star, it is an excellent
  laboratory to study atmospheric escape.  In agreement with the
  theory of atmospheric-loss processes, it is likely that the outer
  planet has a hydrogen-dominated atmosphere. The mass and 
  radius of the inner planet is in agreement with theoretical models 
  predicting an iron core containing $80^{+20}_{-30}$\% of its mass. Such a
  high metal content is surprising, particularly given that the star
  has an ordinary (solar) metal abundance. We discuss various possible
  formation scenarios for this unusual planet.}
   \maketitle


\section{Introduction}
\label{sectI}

In recent years, many planets with masses lower than $\rm
15\,M_{\oplus}$ have been discovered. Surprisingly, these
planets show a great diversity in bulk densities (see, for example,
Hatzes \& Rauer \cite{hatzes15}).  It is obvious that the planets with
the highest densities must be rocky, while those with the lowest
densities must have a large fraction of volatiles such as hydrogen.
Planets of intermediate densities could in principle have many
different compositions, but there is now growing evidence that they
also have rocky cores and less extended hydrogen atmospheres (see, for
example, Chen et al. \cite{chen17}).  A picture has thus emerged in
which the diversity of low-mass exoplanets is explained by the
different size of the hydrogen atmospheres -- some planets have very
extended atmospheres, some have less extended ones, and others do not
have them at all.

Why do some planets have hydrogen atmospheres and others do not? A
crucial element for solving this problem was the result that low-mass
close-in planets ($a\leq0.05$ AU) tend to have high bulk densities. It
is thus unlikely that they have extended hydrogen atmospheres.  These
planets are CoRoT-7b (L\'eger et al. \cite{leger09}), Kepler-10b
(Batalha et al. \cite{batalha11}), Kepler-36b (Carter et al.
\cite{carter12}), Kepler-78b (Sanchis-Ojeda et al. \cite{sanchis13}),
Kepler-93b (Dressing et al. \cite{dressing15}), HD~219134b (Motalebi
et al. \cite{motalebi15}), GJ~1132b, (Berta-Thompson et
al. \cite{berta15}), WASP-47e (Dai et al. \cite{dai15}; Sinukoff et
al. \cite{sinukoff17a}), and HD3167b (Gandolfi et
  al. \cite{gandolfi17}).

This collection of findings led to the
hypothesis that atmospheric escape must play an important role in the
formation and evolution of planets (e.g., Lammer et
al. \cite{lammer14}; Sanchis-Ojeda et al. \cite{sanchis14}; Lundkvist
et al. \cite{lundkvist16}; Cubillos et al. \cite{cubillos17}).  For
example, Cubillos et al. (\cite{cubillos17}) showed that planets with
restricted Jeans escape parameters $\Lambda\leq 20$ cannot retain
hydrogen-dominated atmospheres.  The restricted Jeans escape
  parameter is defined as $\Lambda$=\,$\frac{GM_{\rm pl}m_{\rm
      H}}{k_{\rm B}T_{\rm eq}R_{\rm pl}}$ (Fossati et
  al. \cite{fossati17}; see also the description in Cubillos et
  al. \cite{cubillos17}).

The atmospheres of planets with $\Lambda$ values lower than 20 - 40,
depending on the system parameters, lie in the ``boil-off'' regime
(Owen \& Wu \cite{owen16}; Cubillos et al.\cite{cubillos17}), where
the escape is driven by the atmospheric thermal energy and low
planetary gravity, rather than the high-energy (XUV) stellar flux.
Atmospheric escape can thereby explain why low-mass close-in
planets do not have extended hydrogen atmospheres.

The atmospheric escape rates have been determined for a number of
planets with $\rm M>15\,M_{\oplus}$ by analyzing the profiles of the
Lyman-$\alpha$ lines.  For example, Bourrier et
al. (\cite{bourrier16}) derived an escape rate of $\rm \sim 2.5 \times
10^8\,g\,s^{-1}$ for \object{GJ\,436\,b}. This system is of
  particular interest for atmospheric escape studies because of the
  large hydrogen corona that has been detected around the planet
  (Ehrenreich et al. \cite{ehrenreich15}).  These results clearly
support the idea that atmospheric loss processes are an important
factor in the evolution of low-mass planets. Because the escape rate
depends on the amount of XUV-radiation that a planet receives during
its lifetime, as well as on its mass, it would be ideal to study a system
that has two transiting planets of the same mass but at different
distances from the host star.

Finding such a system and deriving the masses and radii of the planets
is thus important.  Density measurements of planets are not only
important to study atmospheric escape, but also to constrain the
structure of exoplanets, which in turn gives us important clues
as to where and how they formed (Raymond et
al. \cite{raymond13}). In this article, we point out that
  \object{K2-106} is such a system.

Recently, Adams et al. (\cite{adams17}) found that the star
\object{K2-106} (\object{EPIC\,220674823},\object{TYC\,608-458-1}) has
two transiting planets.  The inner planet has an ultra-short period of
$\rm P=0.571308\pm0.00003$\,d (ultra-short period planets have orbital
periods shorter than one day).  Adams et al. (\cite{adams17}) derived
a radius of $\rm R_p=1.46\pm0.14$\,$\rm R_{\oplus}$.  For the outer
planet, these authors derived an orbital period of $\rm
P=13.341245\pm0.0001$\,d, and a radius of $\rm R_p=2.53\pm0.14$\,$\rm
R_{\oplus}$.  This system is particularly interesting because it hosts
an ultra-short period planet that is subject to strong stellar
irradiation and an outer planet at a relatively large distance from
the host star, where the atmospheric escape rate is expected to be
much lower. Within the framework of the KESPRINT collaboration, we
re-derive the stellar fundamental parameters and determine masses,
radii, and densities of the two planets\footnote{ This paper continues
  a series of papers on K2 planet investigations that were previously
  published by two collaborations, ESPRINT and KEST, which have
  recently merged to form the KESPRINT collaboration (see, e.g.,
  Narita et al. \cite{narita17}; Eigm{\"u}ller et
  al. \cite{eigmueller17}).}.  We show that the two planets have
similar masses and not very different densities.  They are thus
particularly interesting for learning more about the formation and
evolution of planets.

\section{Radial velocity measurements}
\label{sectII}

We obtained absolute and relative radial velocities (RVs) using five
different instruments.  The relative RVs were obtained with HDS, PFS,
and FIES and are described in Sect. \ref{sectII1} (results are listed
in Table~\ref{tab:RV1}). The absolute RVs were obtained with HARPS and
HARPS-N, and are described in Sect. \ref{sectII2} (results are listed
in Table~\ref{tab:RV2}).

\subsection{HDS, PFS, and FIES}
\label{sectII1}

{\bf PFS: } Between August 14, 2016, and January 14, 2017, we obtained
13 spectra of \object{K2-106} with the Carnegie Planet Finder
Spectrograph (PFS) (Crane et al. \cite{crane06}, Crane et al.
\cite{crane08}; Crane et al. \cite{crane10}).  PFS is an \'echelle
spectrograph on the 6.5 m Magellan/Clay Telescope at Las Campanas
Observatory in Chile.  It employs an iodine gas cell to superimpose
well-characterized absorption features onto the stellar spectrum. The
iodine absorption lines are used to establish the wavelength scale and
instrumental profile (Crane et al. \cite{crane10}). The detector was
read out in the standard $2 \times 2$ binned mode. Exposure times
ranged from 20 - 40 minutes, giving a signal-to-noise ratio (S/N) of
50 - 140 pixel$^{-1}$ and a resolution of $\lambda/\Delta \lambda \sim
76,000$ in the wavelength range of the iodine absorption lines. An
additional iodine-free spectrum with higher resolution and higher S/N
was obtained to serve as a template spectrum for the Doppler analysis.
The relative RVs were extracted from the spectrum using the techniques
described by Butler et al. (\cite{butler96}). The internal measurement
uncertainties (ranging from 2-4 $\rm m\,s^{-1}$) were determined from
the scatter in the derived RVs based on individual 2~\AA \, chunks of
the spectrum (Butler et al. \cite{butler96}). Since the spectral lines
of the $\rm I_2$-cell are superposed on the stellar spectrum, spectra
taken with the $\rm I_2$-cell were not used to determine the bisectors
(see below in Sect. ~\ref{sectII2}).

{\bf HDS:} We obtained three RV measurements of \object{K2-106} with
the High Dispersion Spectrograph (HDS; Noguchi et
al. \cite{noguchi02}) on the 8.2 m Subaru Telescope. The spectra were
obtained from October 12 to 14, 2016. We used image slicer 2 (Tajitsu
et al. \cite{tajitsu12}), achieving a spectral resolution of
$\lambda/\Delta\lambda\sim 85,000$ and a typical S/N of $70-80$ per
pixel close to the sodium D lines. This instrument is also equipped
with an $\rm I_2$ cell (see Sato et al.  \cite{sato02} for the HDS RV
analysis). As with the PFS, the RVs are measured relative to a
template spectrum taken by the same instrument without the $\rm
I_2$-cell.

{\bf FIES:} We also obtained six RV measurements with the FIbre-fed
Echelle Spectrograph (FIES; Frandsen \& Lindberg \cite{frandsen99};
Telting et al.  \cite{telting14}) on the 2.56 m Nordic Optical
Telescope (NOT) at the Observatorio del Roque de los Muchachos, La
Palma (Spain). The observations were carried out from October 5 to
November 25, 2016, as part of the observing programs 54-205, 54-027,
and 54-211. We used the 1.3$\arcsec$ high-resolution fiber ($\lambda/\Delta
\lambda $\, 67\,000) and set the exposure time to 2700 s, following
the same observing strategy as Gandolfi et al. (\cite{gandolfi15}). 
We traced the RV drift of the instrument by acquiring ThAr spectra with 
long exposures ($\rm T_\mathrm{exp}$\,$\approx$\,35 s) immediately 
before and after each observation. The data were reduced using
standard IRAF and IDL routines. The S/N of the extracted spectra is
about 35 per pixel at 5500~\AA. RVs were derived via multi-order
cross correlations, using the stellar spectrum with the highest
S/N as template.

HIRES RV measurements from the literature: while this article
  was being refereed, we learned that another group had also
  undertaken RV measurements of \object{K2-106} and uploaded their
  article on the preprint server (Sinukoff et
  al. \cite{sinukoff17b}). Their work included 35 relative RV
measurements 
  obtained with Keck-HIRES, which we also included in our analysis.
  
\begin{table}
\caption{RV measurements of K2-106 obtained with PFS$^1$,
  HDS$^2$ , and FIES$^3$.}
\begin{tabular}{l r r l }
\hline
\noalign{\smallskip}
$\rm BJD_{TDB}^4$     & RV$^5$     & $\pm \sigma$ & Instrument \\
-2 450 000  & [\kms] & [\kms]       & \\
\hline
7614.81876      &     0.0054 & 0.0021 & PFS \\ 
7615.82964      &     0.0001 & 0.0022 & PFS \\
7616.82147      &    -0.0012 & 0.0025 & PFS \\
7617.83381      &     0.0155 & 0.0028 & PFS \\
7618.76739      &    -0.0038 & 0.0024 & PFS \\
7621.83249      &     0.0006 & 0.0023 & PFS \\
7623.75032      &    -0.0043 & 0.0029 & PFS \\
7624.73484      &    -0.0151 & 0.0048 & PFS \\
7760.54699      &     0.0000 & 0.0033 & PFS \\
7763.55780      &     0.0035 & 0.0033 & PFS \\
7764.55645      &     0.0144 & 0.0041 & PFS \\
7765.55324      &    -0.0038 & 0.0040 & PFS \\
7767.55174      &     0.0031 & 0.0037 & PFS \\
\hline
7673.98378 &    -0.0095 & 0.0050 & HDS \\ 
7675.04835 &     0.0078 & 0.0053 & HDS \\
7676.01717 &     0.0078 & 0.0051 & HDS \\ 
\hline
7666.65017 &     0.0016 & 0.0050 & FIES \\ 
7668.56785 &     0.0163 & 0.0043 & FIES \\
7669.50586 &     0.0068 & 0.0036 & FIES \\
7683.46006 &     0.0144 & 0.0068 & FIES \\
7684.59951 &     0.0206 & 0.0061 & FIES \\ 
7717.51153 &    -0.0002 & 0.0045 & FIES \\ 
\hline 
\end{tabular}
\label{tab:RV1}
\\
$^1$ RV offset for PFS:  $1.2_{-1.5}^{+1.5}$ $\rm m\,s^{-1}$,
jitter term $3.9_{-1.3}^{+1.7}$ $\rm m\,s^{-1}$. \\
$^2$ RV offset for HDS:  $2.0_{-8.3}^{+8.7}$ $\rm m\,s^{-1}$,
jitter term $10.8_{-7.6}^{+28.7}$ $\rm m\,s^{-1}$. \\
$^3$ RV offset for FIES: $10.2_{-2.4}^{+2.4}$ $\rm m\,s^{-1}$,
jitter term $2.3_{-1.6}^{+3.0}$ $\rm m\,s^{-1}$. \\
RV offset for HIRES: $-2.09_{-0.93}^{+0.91}$ $\rm m\,s^{-1}$, 
jitter term $5.0_{-0.7}^{+0.8}$ $\rm m\,s^{-1}$. \\
$^4$ Barycentric Julian dates are given in barycentric dynamical time. \\
$^5$ Relative RV. \\
\end{table}

\subsection{HARPS-N and HARPS}
\label{sectII2}

We obtained 12 RV measurements with the HARPS-N spectrograph
(Cosentino et al.  \cite{cosentino12}) on the 3.58 m Telescopio
Nazionale Galileo (TNG) at La Palma in programs CAT16B-61, A34TAC\_10,
A34TAC\_44, and 20 RVs with the HARPS spectrograph (Mayor
\cite{mayor03}) on the 3.6 m ESO telescope at La Silla in program
098.C-0860.  The HARPS-N spectra were obtained from October 30 2016 to
January 28 2017, and the HARPS spectra from October 25 to November 27
2016.  Both spectrographs have a resolving power
$\lambda/\Delta\lambda \sim 115\,000$. HARPS-N covers the wavelength
region from 3780\,\AA\, to 6910\,\AA, and HARPS from 3830\,\AA\, to
6900\,\AA.  All calibration frames were taken using the standard
procedures for these instruments.  The spectra were reduced and
extracted using the dedicated HARPS/HARPS-N pipelines.  The RVs were
determined by using a cross-correlation method with a numerical mask
that corresponds to a G2 star (Baranne et al. \cite{baranne96}; Pepe
et al. \cite{pepe02}).  The RV measurements were obtained by fitting a
Gaussian function to the average cross-correlation function (CCF).
The data reduction pipelines for both instruments also provide the
absolute RV, and the bisector span. Because of the high resolution of
the HARPS spectrographs, these spectra are particularly useful for the
bisector analysis.  We extracted the S-index and $\rm log\,R'_{HK}$
activity indicators from the HARPS and HARPS-N spectra. The measurements
obtained with HARPS-N, and HARPS are listed in Table~\ref{tab:RV2}.

\begin{table*}
\caption{RV measurements K2-106 obtained with HARPS-N$^1$ and HARPS$^2$.}
\begin{tabular}{l r r l l l l l l }
\hline
\noalign{\smallskip}
$\rm BJD_{TDB}^3$     & RV$^4$     & $\pm \sigma$ & Instrument & FWHM   & BIS    & Ca\,II-S-index & $\rm log\,R'_{HK}$ & S/N \\
-2 450 000  & [\kms] & [\kms]       &            & [\kms] & [\kms] &        & & \\
\hline
7692.37945 &   -15.7430 & 0.0034 & HARPS-N &  6.8266 & -0.0452 & $0.164\pm 0.013$ & $-4.94\pm 0.08$ & $27.3\pm 1.2$ \\ 
7692.44910 &   -15.7332 & 0.0028 & HARPS-N &  6.8274 & -0.0461 & $0.166\pm 0.010$ & $-4.93\pm 0.06$ & $31.6\pm 1.1$ \\
7692.53001 &   -15.7332 & 0.0031 & HARPS-N &  6.8373 & -0.0491 & $0.159\pm 0.011$ & $-4.97\pm 0.07$ & $30.8\pm 1.1$ \\
7692.60284 &   -15.7323 & 0.0017 & HARPS-N &  6.8316 & -0.0442 & $0.154\pm 0.005$ & $-5.01\pm 0.04$ & $49.0\pm 1.5$ \\
7693.37242 &   -15.7358 & 0.0020 & HARPS-N &  6.8255 & -0.0266 & $0.138\pm 0.011$ & $-5.14\pm 0.10$ & $32.7\pm 1.2$ \\
7693.45891 &   -15.7400 & 0.0040 & HARPS-N &  6.8420 & -0.0367 & $0.154\pm 0.016$ & $-5.01\pm 0.11$ & $25.8\pm 1.1$ \\
7693.52649 &   -15.7428 & 0.0033 & HARPS-N &  6.8219 & -0.0340 & $0.169\pm 0.013$ & $-4.91\pm 0.07$ & $29.4\pm 1.1$ \\
7693.62313 &   -15.7309 & 0.0040 & HARPS-N &  6.8275 & -0.0195 & $0.194\pm 0.017$ & $-4.79\pm 0.07$ & $25.9\pm 1.8$ \\
7694.37831 &   -15.7309 & 0.0027 & HARPS-N &  6.8147 & -0.0402 & $0.146\pm 0.009$ & $-5.07\pm 0.08$ & $33.2\pm 1.2$ \\
7694.46390 &   -15.7341 & 0.0033 & HARPS-N &  6.8333 & -0.0517 & $0.150\pm 0.012$ & $-5.04\pm 0.09$ & $29.0\pm 1.0$ \\
7694.53229 &   -15.7372 & 0.0025 & HARPS-N &  6.8133 & -0.0467 & $0.147\pm 0.008$ & $-5.06\pm 0.07$ & $35.4\pm 1.1$ \\
7782.37246 &   -15.7278 & 0.0032 & HARPS-N &  6.8373 & -0.0370 & $0.148\pm 0.011$ & $-5.05\pm 0.09$ & $31.4\pm 1.7$ \\
\hline 
7686.68137 &   -15.7431 & 0.0046 & HARPS &   6.8855 & -0.0199 & $0.144\pm0.023$ & $  -5.09\pm 0.19$ & $  23.1\pm 1.3$ \\  
7688.59984 &   -15.7245 & 0.0032 & HARPS &   6.8922 & -0.0441 & $0.151\pm0.014$ & $  -5.03\pm 0.11$ & $  31.0\pm 1.4$ \\
7689.61435 &   -15.7419 & 0.0030 & HARPS &   6.8933 & -0.0249 & $0.123\pm0.013$ & $  -5.32\pm 0.19$ & $  32.4\pm 1.3$ \\
7689.66403 &   -15.7353 & 0.0024 & HARPS &   6.9075 & -0.0310 & $0.172\pm0.010$ & $  -4.90\pm 0.06$ & $  38.6\pm 1.3$ \\
7690.63488 &   -15.7407 & 0.0032 & HARPS &   6.8968 & -0.0264 & $0.146\pm0.016$ & $  -5.07\pm 0.13$ & $  30.9\pm 1.3$ \\
7690.70790 &   -15.7422 & 0.0031 & HARPS &   6.9103 & -0.0327 & $0.187\pm0.015$ & $  -4.87\pm 0.08$ & $  32.2\pm 1.4$ \\
7691.58078 &   -15.7225 & 0.0024 & HARPS &   6.8934 & -0.0349 & $0.134\pm0.010$ & $  -5.18\pm 0.10$ & $  38.9\pm 1.4$ \\
7691.69428 &   -15.7339 & 0.0027 & HARPS &   6.9038 & -0.0343 & $0.118\pm0.013$ & $  -5.60\pm 0.35$ & $  36.1\pm 1.4$ \\
7694.63340 &   -15.7395 & 0.0033 & HARPS &   6.9107 & -0.0278 & $0.230\pm0.016$ & $  -4.66\pm 0.05$ & $  30.5\pm 1.3$ \\
7694.70792 &   -15.7359 & 0.0035 & HARPS &   6.9018 & -0.0079 & $0.126\pm0.018$ & $  -5.43\pm 0.34$ & $  29.5\pm 1.5$ \\
7695.59904 &   -15.7294 & 0.0044 & HARPS &   6.9131 & -0.0368 & $0.184\pm0.020$ & $  -4.84\pm 0.10$ & $  24.2\pm 1.3$ \\
7695.68251 &   -15.7343 & 0.0042 & HARPS &   6.9365 & -0.0160 & $0.198\pm0.020$ & $  -4.82\pm 0.09$ & $  25.3\pm 1.4$ \\
7696.56824 &   -15.7246 & 0.0034 & HARPS &   6.9029 & -0.0352 & $0.081\pm0.015$ & $^6$              & $  29.1\pm 1.4$ \\
7696.64271 &   -15.7293 & 0.0029 & HARPS &   6.8900 & -0.0354 & $0.158\pm0.013$ & $  -5.05\pm 0.10$ & $  33.9\pm 1.3$ \\
7697.58844 &   -15.7326 & 0.0035 & HARPS &   6.9034 & -0.0244 & $0.141\pm0.017$ & $  -5.11\pm 0.15$ & $  29.0\pm 1.3$ \\
7697.67020 &   -15.7336 & 0.0030 & HARPS &   6.9030 & -0.0337 & $0.131\pm0.015$ & $  -5.21\pm 0.17$ & $  32.9\pm 1.4$ \\
7717.53993 &   -15.7363 & 0.0036 & HARPS &   6.9077 & -0.0087 & $0.177\pm0.016$ & $  -4.93\pm 0.10$ & $  27.6\pm 1.3$ \\
7717.60935$^5$ & -15.7502 & 0.0056 & HARPS & 6.8841 & -0.0370 & $0.202\pm0.034$ & $  -4.76\pm 0.14$ & $  20.3\pm 1.4$ \\
7719.53421 &   -15.7280 & 0.0028 & HARPS &   6.9022 & -0.0328 & $0.131\pm0.014$ & $  -5.22\pm 0.16$ & $  34.4\pm 1.3$ \\
7719.60121 &   -15.7235 & 0.0033 & HARPS &   6.9028 & -0.0428 & $0.165\pm0.017$ & $  -4.94\pm 0.10$ & $  30.8\pm 1.3$ \\
\hline
\end{tabular}
\label{tab:RV2}
\\
$^1$ Systemic RV of HARPS-N: $-15735.77_{-1.18}^{+1.20}$ $\rm m\,s^{-1}$, 
jitter term $1.9_{-1.2}^{+1.5}$ $\rm m\,s^{-1}$. \\
$^2$ Systemic RV of HARPS: $-15732.70_{-0.92}^{+0.90}$  $\rm m\,s^{-1}$, 
jitter term $4.9_{-0.65}^{+0.76}$ $\rm m\,s^{-1}$. \\
$^3$ Barycentric Julian dates are given in barycentric dynamical time. \\
$^4$ Absolute RV. \\
$^5$ Spectrum with very low S/N, not used for the fit. \\
$^6$ Value could not be obtained.\\
\end{table*}

\section{Combined analysis and properties of the host star and the planets}
\label{sectIII}

\subsection{Properties of the host star}
\label{sectIII.1}

\object{K2-106} (\object{EPIC\,220674823},\object{TYC\,608-458-1}) is
a G5V star with V=12.10, located at RA: $00^h$ $52^m$ $19.147^s$, DEC:
$+10\degr$ $47\arcmin$ $40.92\arcsec$ 
($\rm l=123.2840^o$ $\rm b=-52.0764^o$).  The photospheric parameters, that is,
effective temperature $\rm T_{\rm eff}$, surface gravity $\rm log(g)$,
metal content [M/H], and projected rotation velocity $\rm v \sin i$,
were determined spectroscopically by Adams et al. (\cite{adams17})
along with the stellar mass and radius. The authors used three spectra
with S/N between 30 and 60 per resolution element at 5650~\AA\,
obtained with the Tull Coud\'e spectrograph of the 2.7 m telescope at
the McDonald Observatory. Although the resolution was not specified in the
article, it is presumably $\lambda/\Delta \lambda $\,$\sim$ \,60\,000.

Since our results depend critically on the stellar parameters, we
decided to carry out our own spectral analysis. We used the coadded
HARPS-N and HARPS spectra, which have an S/N of about 240 at
5650~\AA\ per resolution element and a resolving power of
$\lambda/\Delta \lambda $\,=\,115\,000. Our analysis follows the
method outlined by Johnson et al. (\cite{johnson16}). We used SME
version 4.43 (Valenti \& Piskunov \cite{valenti96}; Valenti \& Fischer
\cite{valenti05}) and a grid of the ATLAS12 model atmospheres (Kurucz
\cite{kurucz13}) to fit spectral features that are sensitive to
different photospheric parameters. We adopted the calibration
equations of Bruntt et al. (\cite{bruntt10}) to estimate the
microturbulent velocity and fit many isolated and unblended metal
lines to determine the projected rotation velocity ($\rm
v \sin i$). We derived an effective temperature $\rm
T_{\rm eff}=5470\pm30$\,K , surface gravity $\rm log(g)=4.53\pm0.08$
(cgs), and iron content of $\rm [Fe/H]=-0.025\pm0.050$ dex. We also
derived the abundances of other elements (see Table~\ref{tab:star}).

We obtained the stellar mass and radius using the \texttt{PARSEC}
  model isochrones along with the online interface \footnote{Available
    at http://stev.oapd.inaf.it/cgi-bin/param\_1.3.} for Bayesian
  estimation of the stellar parameters from da Silva et
  al. (\cite{dadilva06}). For \object{K2-106} we derive a mass of
  $M_\mathrm{*}$=0.945$\pm$0.063\,$M_\odot$ and radius of
  $R_\mathrm{*}$=0.869$\pm$0.088\,$R_\odot$ (Table~\ref{tab:star}).
  These values can be compared with those derived by Adams et
  al. (\cite{adams17}), who derived $0.93\pm0.01$ $\rm M_{*}[M\odot]$
  and $0.83\pm0.04$ $\rm R_{*}[R\odot]$, respectively. Although
  these values are the same within 1$\sigma$ , it is interesting to
  note that values derived by Adams et al. (\cite{adams17}) lead to
  higher densities for the planets.

We can test and verify the spectroscopic determination using the Gaia
parallax (3.96$\pm$0.78 mas; d=253$\pm$50 pc; Gaia collaboration et
al.  \cite{gaia2016a}; Gaia collaboration et al. \cite{gaia2016b};
Lindegren et al. \cite{lindegren16}). The basic idea is that the
radius and mass of the star can be determined from the luminosity,
$\rm T_{\rm eff}$ , and the iron abundance without using the spectroscopic
determination of the surface gravity, which is notoriously difficult
to measure.  The luminosity is derived from the apparent magnitudes
and the parallax. An advantage of this method is that the stellar
parameters will be determined with a much higher accuracy using
forthcoming data from Gaia. However, to use this method, we also
have to know to which degree the apparent brightness of the star is
affected by extinction. Following the method described by Gandolfi et
al. (\cite{gandolfi08}), we derived the interstellar extinction
$A_\mathrm{v}$ by fitting the spectral energy distribution of the star
to synthetic colors extracted from the NextGen model spectrum with
the same photospheric parameters as the star. We find an extinction of
$A_\mathrm{v}$=0.1$\pm$0.1\,mag, as expected given the relatively
nearby location (see below) and high galactic latitude of the
star. The effect from the extinction is negligible, and
we determined the radius and mass of the star using the \texttt{PARSEC}
model isochrones. Using this method, we derive a stellar mass of
$M_\mathrm{*}$=0.902$\pm$0.027\,$M_\odot$ and radius of
$R_\mathrm{*}$=0.882$\pm$0.050\,$R_\odot$ (Table~\ref{tab:star}),
which implies a surface gravity of $\rm log(g)=4.474\pm0.053$ (cgs).
The mass and radius of the star derived by this method is again
  the same within  1$\sigma$  as our spectroscopic determination and
  the values derived by Adams et al. (\cite{adams17}).

For the purposes of the present paper, we used our stellar
parameter estimates because they are based on spectra with higher
resolution and S/N than those used in previous works. However,
  for completeness, we also give the masses and radii for the two
  planets derived using stellar parameters from Adams et
  al. (\cite{adams17}).

\begin{table*}
\caption{Properties of the host star.}
\begin{tabular}{l c c c }
\hline
\noalign{\smallskip}
                      & Adams et al. (\cite{adams17}) & Gaia
and $\rm T_{\rm eff}$  & this work$^1$ \\
\hline
$\rm M_{*}[M\odot]$ & $0.93\pm0.01$   & $0.902\pm0.027^2$ & $0.945\pm0.063$ \\
$\rm R_{*}[R\odot]$ & $0.83\pm0.04$   & $0.882\pm0.050^2$ & $0.869\pm0.088$ \\
$\rm T_{\rm eff}$ [K]      & $5590\pm51$     &    $\ldots$              & $5470\pm30$   \\
$\rm log(g)$           & $4.56\pm0.09$   & $4.474\pm0.053^2$ & $4.53\pm0.08$ \\
$\rm [Fe/H]$           & $0.025\pm0.020$ &                 & $-0.025\pm0.05$ \\
$\rm [Si/H]$           & $\ldots$ & $\ldots$  & $-0.05\pm0.05$ \\
$\rm [Ca/H]$           & $\ldots$  & $\ldots$ & $+0.08\pm0.05$ \\
$\rm [Ni/H]$           & $\ldots$ & $\ldots$ & $-0.02\pm0.05$ \\
$\rm [Na/H]$           & $\ldots$ & $\ldots$ & $+0.05\pm0.05$ \\
$\rm v \sin i$ [$\rm km\,s^{-1}$] & $\ldots$ & $\ldots$ & $2.8\pm0.35$ \\ 
$\rm v_{macro}$ [$\rm km\,s^{-1}$] & $\ldots$ & $\ldots$ & $1.7\pm0.35$ \\
$\rm v_{micro}$ [$\rm km\,s^{-1}$] & $\ldots$ & $\ldots$ & $0.9\pm0.1^3$ \\
$\rm A_\mathrm{v}$ [mag] & $\ldots$ & $\ldots$ & $0.1\pm0.1$\ \\ 
distance [pc]           & $\ldots$ & $253\pm50$ & $\ldots$ \\
\hline
\end{tabular}
\label{tab:star}
\\
$^1$ Spectroscopic determination as derived from the HARPS and HARPS-N spectra. \\ 
$^2$ Derived using $\rm T_{\rm eff}$, $\rm [Fe/H]$ from HARPS
and HARPS-N, and the Gaia parallax in Sect.~\ref{sectIII.1}.\\
$^3$ Using the empirical formula from Bruntt et al. (\cite{bruntt10}). \\
\end{table*}

\subsection{Activity of the host star}
\label{sectIII.2}

\begin{figure}
\includegraphics[height=.37\textheight,width=.22\textheight,angle=-90.0]{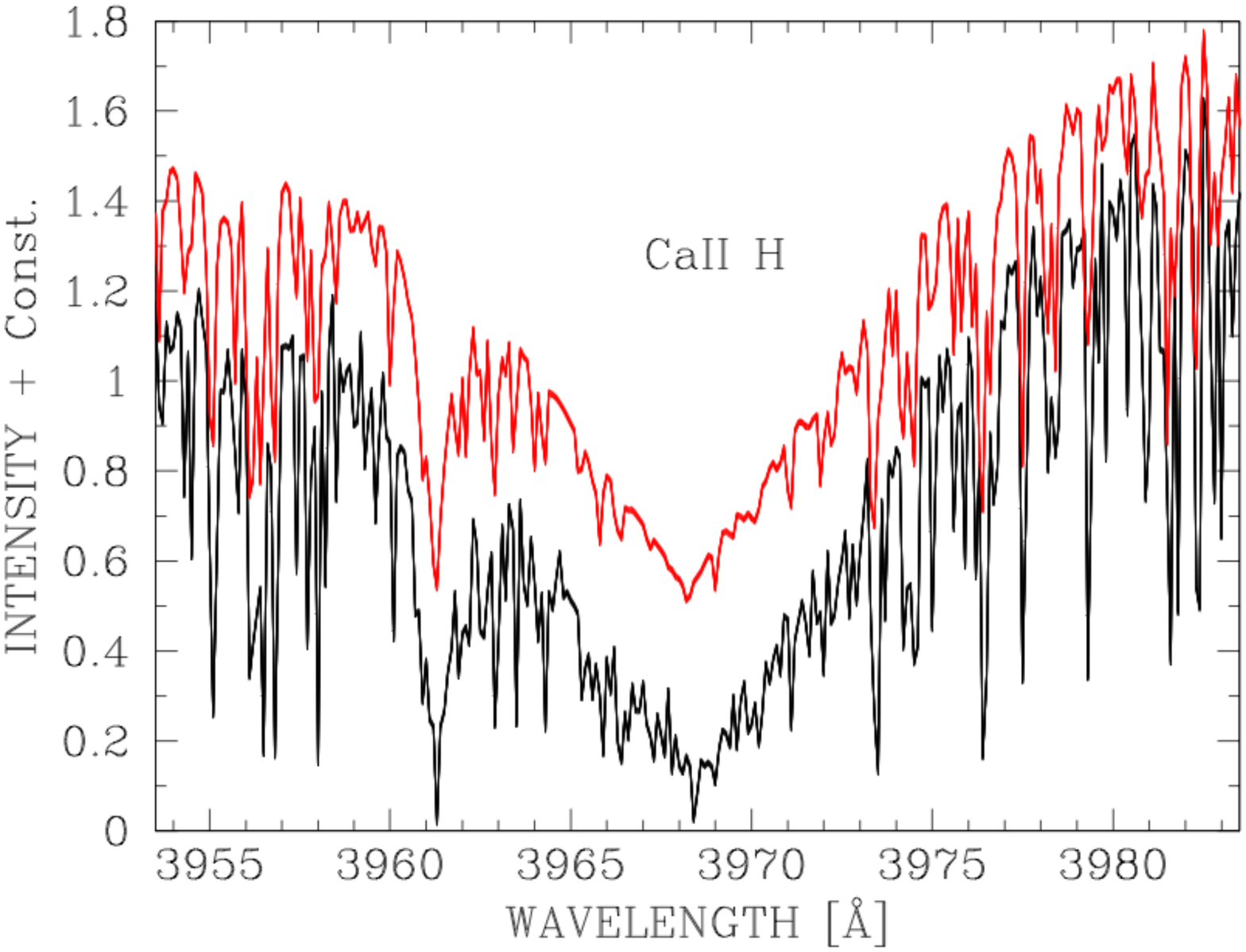}
\caption{Averaged HARPS spectrum of \object{K2-106} (black)
  in the Ca\,II\,H line together with a solar spectrum (red).}
\label{EPICCaIIHK}
\end{figure}

\begin{figure}
\includegraphics[height=.22\textheight,angle=0.0]{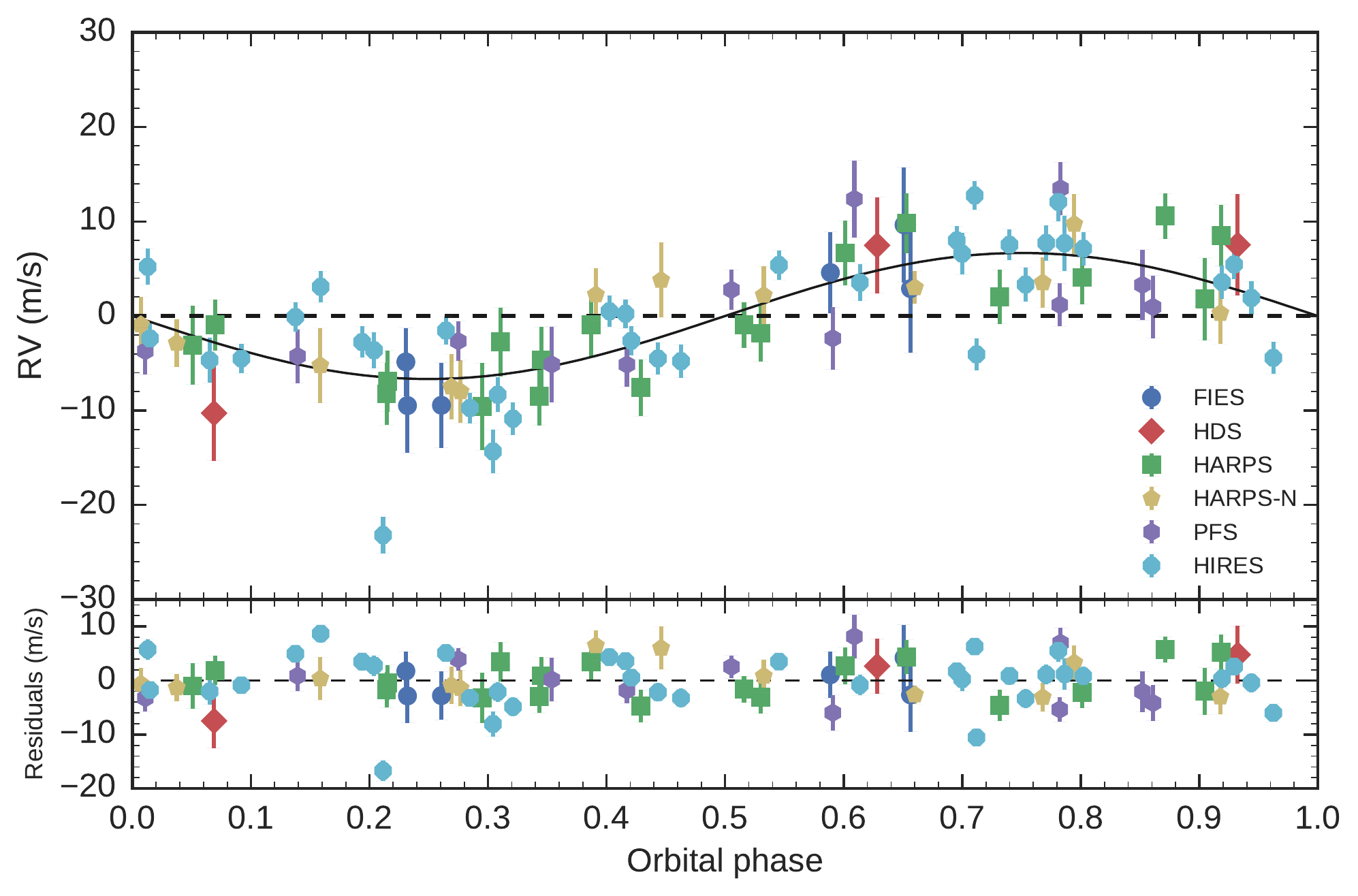}
\caption{Phase-folded RV curve of \object{K2-106\,b}
  after removing the signal from the outer planet.}
\label{EPICbRV}
\end{figure}

\begin{figure}
\includegraphics[height=.22\textheight,angle=0.0]{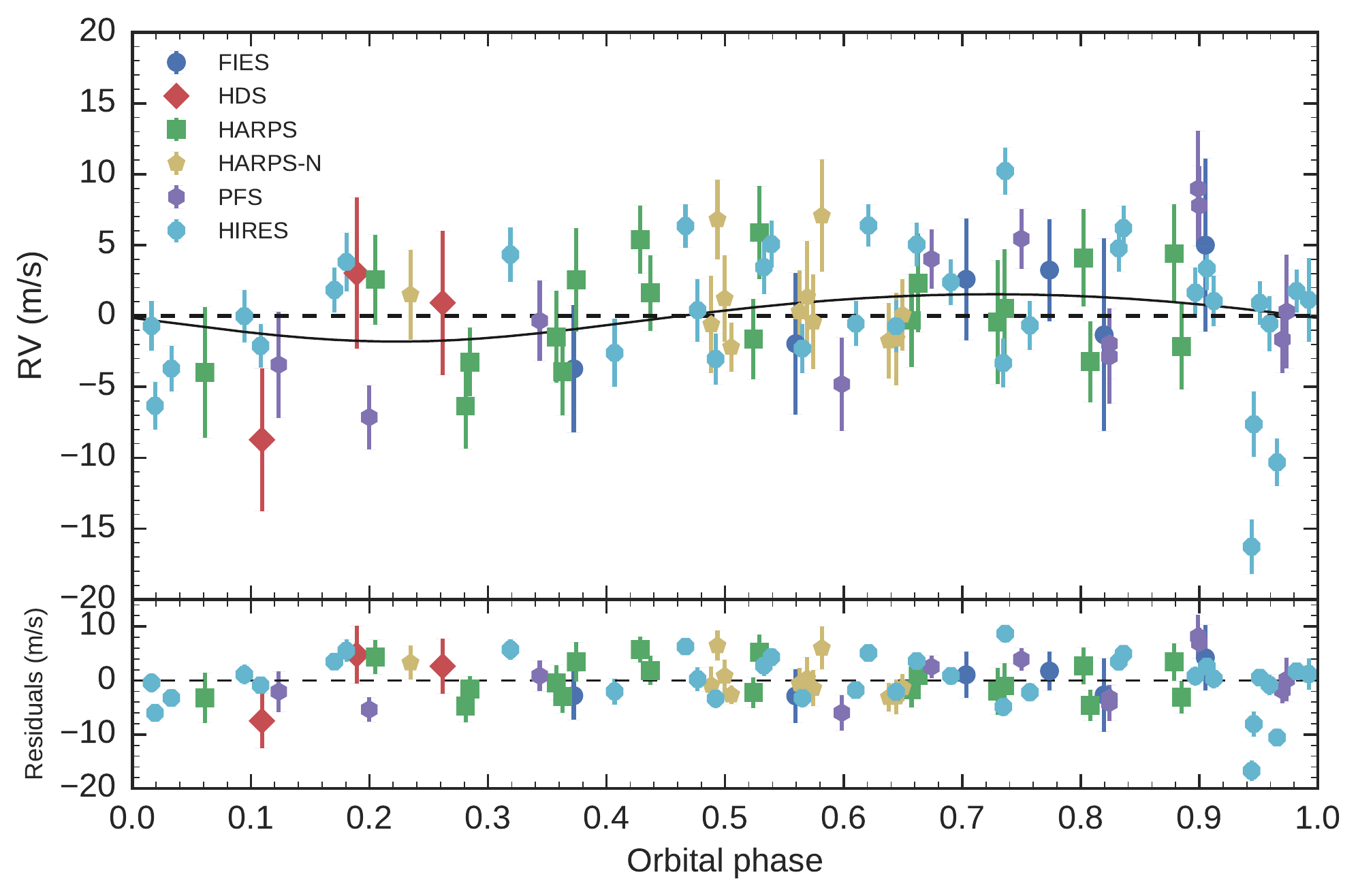}
\caption{Phase-folded RV curve of \object{K2-106\,c}
  after removing the signal from the inner planet.}
\label{EPICcRV}
\end{figure}

Before discussing the RV signals of the planets, we need to know
whether stellar activity affects the RV measurements or the light
curves.  From the HARPS and HARPS-N spectra we derive an average
chromospheric activity index $\rm log\,R'_{HK}=-5.04\pm0.19$
(Table~\ref{tab:RV2}). As pointed out by Saar (\cite{saar06}), the
minimum chromospheric activity of stars with solar metallicity is
about $\rm log\,R'_{HK}=-5.08$. Since we do not see any emission
component in the Ca\,II\,H\&K lines (Fig.~\ref{EPICCaIIHK}) either, we
conclude that the star is very inactive, in agreement with its slow
rotation of $\rm v \sin i=2.8\pm0.35$ $\rm km\,s^{-1}$.  This does not
imply, however, that there is no RV jitter caused by stellar activity.
Lanza et al. (\cite{lanza16}) showed that the amplitude of the
long-term RV variation of the Sun in the time from 2006 to 2014 was
$\rm 4.98\pm1.44\, m\,s^{-1}$. At the maximum of the solar activity,
the amplitude of the RV variations can be as high as $\rm
8\,m\,s^{-1}$ (Meunier et al \cite{meunier10a}).  The scatter of the
RV measurements shown in Figs.~\ref{EPICbRV} and \ref{EPICcRV} appears
to be dominated by the photon noise of the spectra, not by stellar
activity, which is consistent with the result that this star is as
inactive as the Sun.

Although the orbital periods of the planets are already known from the
transit light curve, it is nevertheless useful to perform a period
search to investigate whether stellar activity could systematically
change the inferred RV amplitudes of the planets, or whether it merely
adds random noise to the data.  Since the RV variations induced by
activity on the Sun are correlated with the $\rm log\,R'_{HK}$-index
(Meunier et al \cite{meunier10b}), we calculated the Lomb-Scargle
diagram for the stellar $\rm log\,R'_{HK}$ and the bisector span.  The
are no significant peaks (false-alarm probability lower than 1\%) at
the orbital periods of the planets, which means that the observed RV
variations are not induced by stellar activity.

\begin{figure}
\includegraphics[height=.22\textheight,angle=0.0]{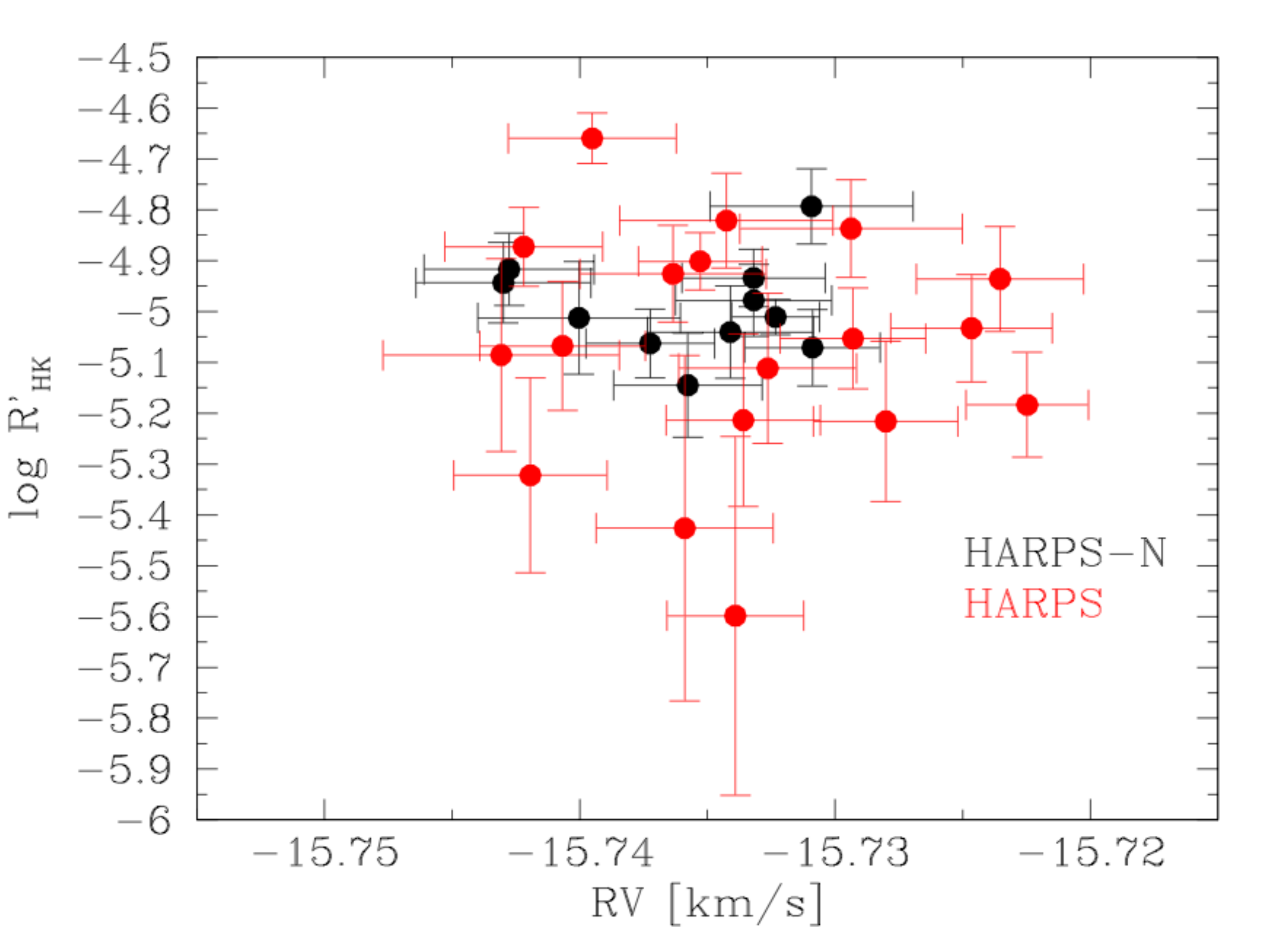}
\caption{Same as Fig.~\ref{EPICbis}, but for the chromospheric
  activity index $\rm log\,R'_{HK}$.  There is again no correlation
  between the two.}
\label{EPIClogRhk}
\end{figure}

\begin{figure}
\includegraphics[height=.22\textheight,angle=0.0]{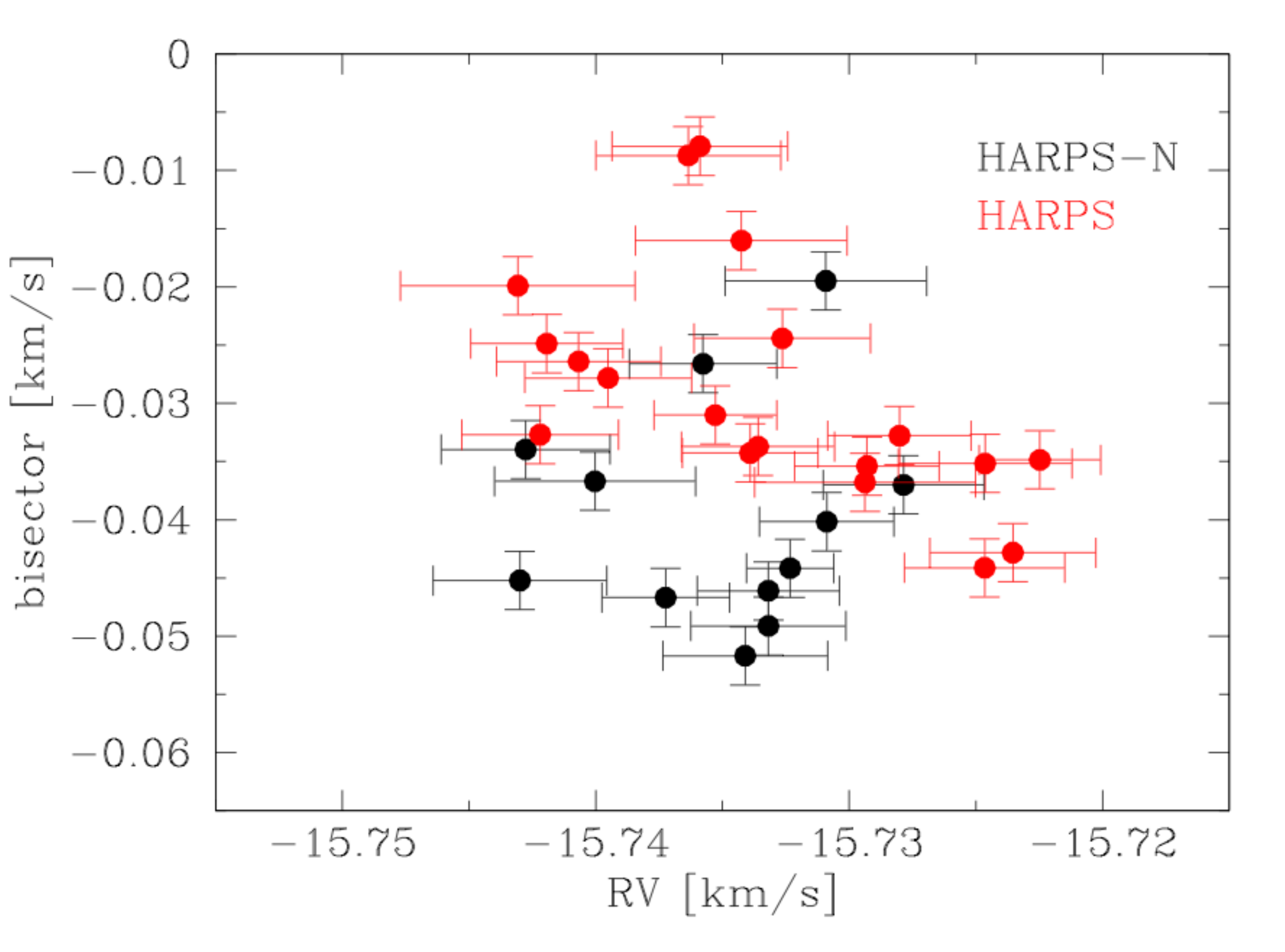}
\caption{Bisector span versus the RV for \object{K2-106}.  There is no
  correlation between the two, indicating that stellar activity or
  hypothetical background binaries probably do not affect the
  derived RV amplitudes for the two planets.}
\label{EPICbis}
\end{figure}

In Figs.~\ref{EPIClogRhk} and Fig.~\ref{EPICbis} we plot $\rm
log\,R'_{HK}$ and the bisector span against RV. The correlation
coefficient between $\rm log\,R'_{HK}$ and the RV is $-0.07\pm0.10$
and the correlation between the bisector span and the RV is
$-0.26\pm0.23$.  This means that there are no significant correlations
between the activity indicators and the RV variations, suggesting that
stellar activity does not significantly bias the RV amplitudes ($K$
values).  Although the activity of the star is low, we nevertheless
include a jitter term in the analysis.  The jitter terms and the
systemic velocities are given in Tables~\ref{tab:RV1}
and~\ref{tab:RV2}.  However, we also quote the results obtained
without using the jitter term, to discuss whether the inclusion of a
jitter term makes any significant difference.

\subsection{Multi-planet joint analysis}
\label{sectIII.3}

We performed a joint analysis of the K2 light curve and RV data of
\object{K2-106}. We used the K2 photometry provided by Vanderburg \&
Johnson (\cite{vanderburg14}), and detrended and cleaned the transit
light curves using the code \texttt{exotrending}\footnote{Available at
  {https://github.com/oscaribv/exotrending} (Barrag{\'a}n \& Gandolfi
  \cite{barragan17})}.  For each transit light curve,
\texttt{exotrending} fits a second-order polynomial to the
out-of-transit data. The fitted segments includes 4 and 12 hours of
out-of-transit data centered around each transit of the inner and
outer planet, respectively. The code removes outliers using a
3$\sigma$ clipping algorithm applied to the residuals to the
preliminary best-fitting transit model derived using the equations
from Mandel \& Agol (\cite{mandel02}), coupled to a nonlinear
least-squares fitting procedure.

The multi-planet joint analysis was made with the code
\texttt{pyaneti} (Barrag\'an et al. \cite{barragan17}). This code
explores the parameter space with a Markov chain Monte Carlo algorithm
and generates a posterior distribution for each parameter. The transit
fits are made using a Mandel \& Agol (\cite{mandel02}) model, while we
used Keplerian orbits to model the RV measurements.  The likelihood
and fitted parameters are the same as in Barrag{\'a}n et al.
(\cite{barragan16}).  For each planet, the fitted parameters are
listed in Table~\ref{tab:values}.  Briefly, they are 1) the time of
first transit $T_{0}$; 2) the orbital period $P$; 3) $\sqrt{e} \sin
\omega_\star$ and \emph{4})$ \sqrt{e} \cos \omega_\star$, where $e$ is
the eccentricity and $\omega_{\star}$ is the argument of periapsis of
the star; 5) the impact parameter $b$, defined as $\cos i\, [1 - e^2]
/ [ R_\star (1 + e \sin \omega_\mathrm{p})]$, where $i$ is the orbital
inclination with respect to the line of sight, $R_\star$ is the
stellar radius, and $\omega_\mathrm{p}$ is the argument of periapsis
of the planet; 6) the scaled semi-major axis $a / R_{\star}$; 7) the
planet-to-star radius ratio $R_{\rm p} / R_{\star}$; 8) the RV
semi-amplitude variation $K$; and 9) the systemic velocities
$\gamma_j$ for each instrument $j$.  The code also fits for the
limb-darkening coefficients $u_1$ and $u_2$ using the parameterization
$q_1$ and $q_2$ proposed by Kipping et al.  (\cite{kipping13}).
Table~\ref{tab:values} reports also the derived quantities, namely,
the planetary mass $M_{\rm p}$ and radius $R_{\rm p}$, bulk density
$\rho_{\rm p}$, surface gravity $g_{\rm p}$, equilibrium temperature
$T_{\rm eq}$ (assuming zero albedo), as well as the transit duration
$\tau_{14}$ and the ingress/egress duration $\tau_{12} = \tau_{34}$.

The long-cadence data give a slightly distorted view of the actual
transit shape.  To take this into account, we followed the procedure
described by Kipping et al.  (\cite{kipping10}). We subdivided each
time stamp into ten points, calculated the theoretical flux for each
point, and then performed an average before comparing to the data. We
set uniform priors for the following parameters within the ranges
$T_{0,b} = [2457394.00,2457394.02 ]~\mathrm{d}$, $T_{0,c} =
[2457405.69 , 2457405.77]~\mathrm{d}$, $P_{b} = [0.5710 ,
  0.5716]~\mathrm{d}$, $P_{c} = [13.33, 13.35]~\mathrm{d}$, $b_i =
[0,1]$, $K_i = [0,100]~\mathrm{m\,s^{-1}}$, and $R_{{\rm p},i} / R_\star =
[0,0.1]$. For circular orbits the parameters $ \sqrt{e_i} \sin
\omega_i$, $ \sqrt{e_i} \cos \omega_i$ were fixed to 0, whereas for
eccentric orbits the priors for the two eccentricity parameters were
uniform between -1 and 1, taking into account that $e < 1$. For the the
limb-darkening coefficients $u_1$ and $u_2$, we adopted Gaussian
priors centered at the values given by Claret \& Bloemen
(\cite{claret11}) with conservative error bars of $0.1$. For the
scaled semi-major axis, we used Kepler's third law to set Gaussian
priors based on the stellar mass and radius as derived in
Sect.~\ref{sectIII.1}.

The parameter space was explored using 500 independent Markov
chains. Once the chains converged to a solution, we ran 25,000
additional iterations with a thin factor of 50. This produced a
posterior distribution of 250,000 independent points for each
parameter. The final parameters and their corresponding error bars
were defined by the median and the 68\% levels of the credible
interval of the posterior distribution.

Given the very short orbital period, we assumed a circular orbit for
\object{K2-106\,b}, but included eccentricity orbit in the case
  of \object{K2-106\,c}. Using the full analysis, all data, and the
  jitter term, we find $\rm e_c=0.18_{-0.12}^{+0.15}$ for
  \object{K2-106\,c}.  Figure~\ref{EPICcRV} shows the phase-folded
  RV curve and the orbit with an eccentricity of 0.18.

There are in principle four possibilities for obtaining the mass of
the two planets: We can use just our data, or we can also include the
data taken by Sinukoff et al. (\cite{sinukoff17b}), and we can perform
the analysis with and without the jitter term. The K-amplitudes using
our data without the jitter term are $K_b=6.25\pm0.63$ $m\,s^{-1}$ and
$K_c=2.38\pm0.80$ $m\,s^{-1}$. With the jitter terms they are
$K_b=6.39\pm0.85$ $m\,s^{-1}$, and $K_c=1.76\pm1.0$ $m\,s^{-1}$. The
effect of the jitter term is thus small, as these values are the same
within 1$\sigma$ .  When we include the measurements taken by Sinukoff
et al. (\cite{sinukoff17b}) and the jitter term, we find
$K_b=6.67\pm0.69$ $m\,s^{-1}$, $K_c=1.67_{-0.88}^{+0.99}$ $m\,s^{-1}$
The inclusion of a jitter term and the data from Sinukoff et
al. (\cite{sinukoff17b}) thus does not change the results
significantly, but the accuracy of the mass determination increases
slightly when we include the data from Sinukoff et
al. (\cite{sinukoff17b}).  In the following we use the values obtained
with the jitter term and including the data taken by Sinukoff et
al. (\cite{sinukoff17b}).

Using the masses and orbital parameters of the two planets, we estimated
the expected transit-time-variations (TTVs) induced by the
gravitational mutual interactions between the two objects.  Because
the two planets are not in resonance, the interaction between the
two planets is very small. The resulting TTVs are too small to be
detected using {\it Kepler} long-cadence data.

\subsection{Radii, masses, and densities of the planets}

\begin{figure}
\includegraphics[height=.22\textheight,angle=0.0]{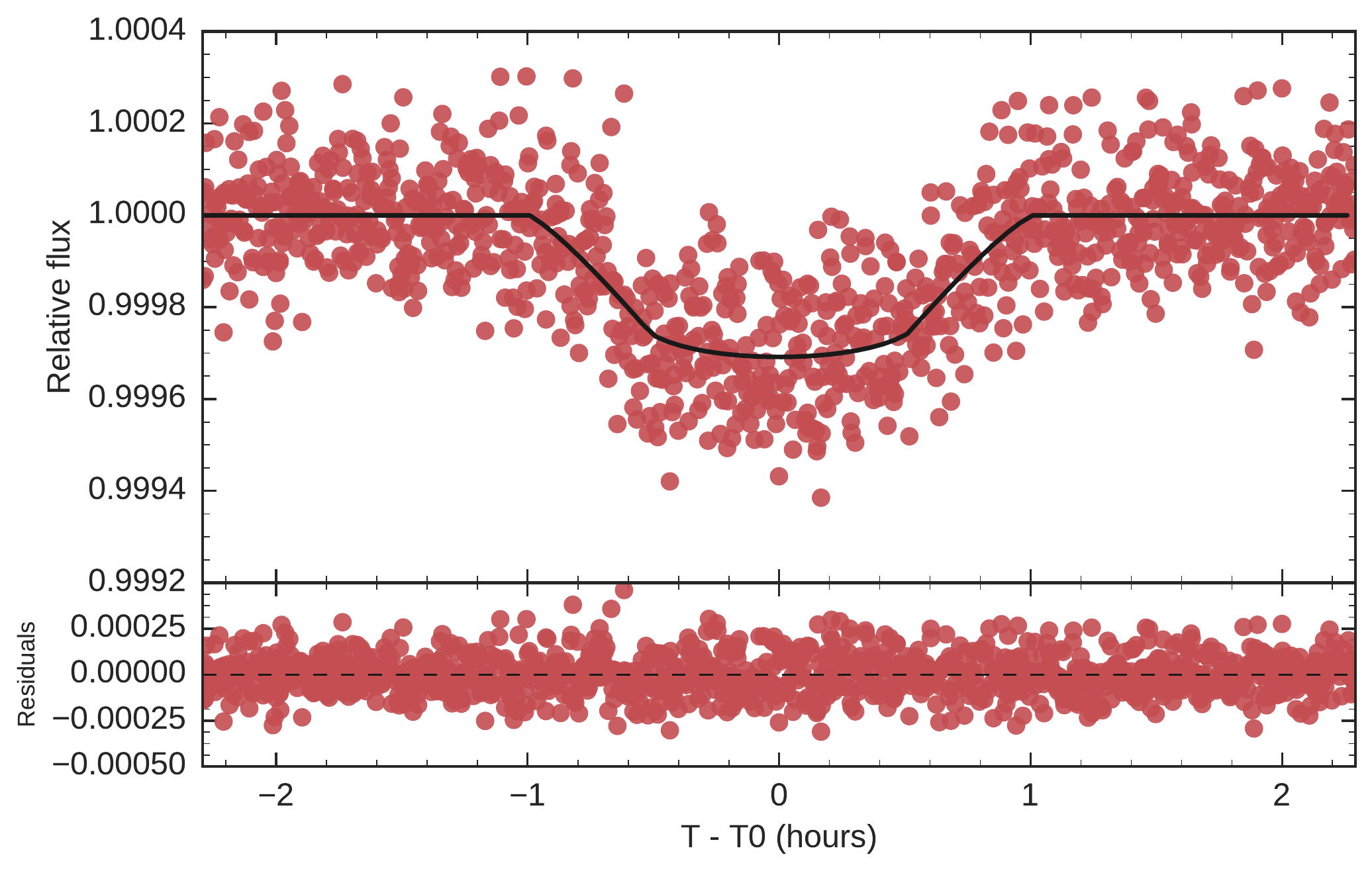}
\caption{Best-fit light curves of the planet \object{K2-106\,b}.
  The light curve has been folded using the orbital period of the planet 
  (Table~\ref{tab:values}).}
\label{EPICbTR}
\end{figure}

\begin{figure}
\includegraphics[height=.22\textheight,angle=0.0]{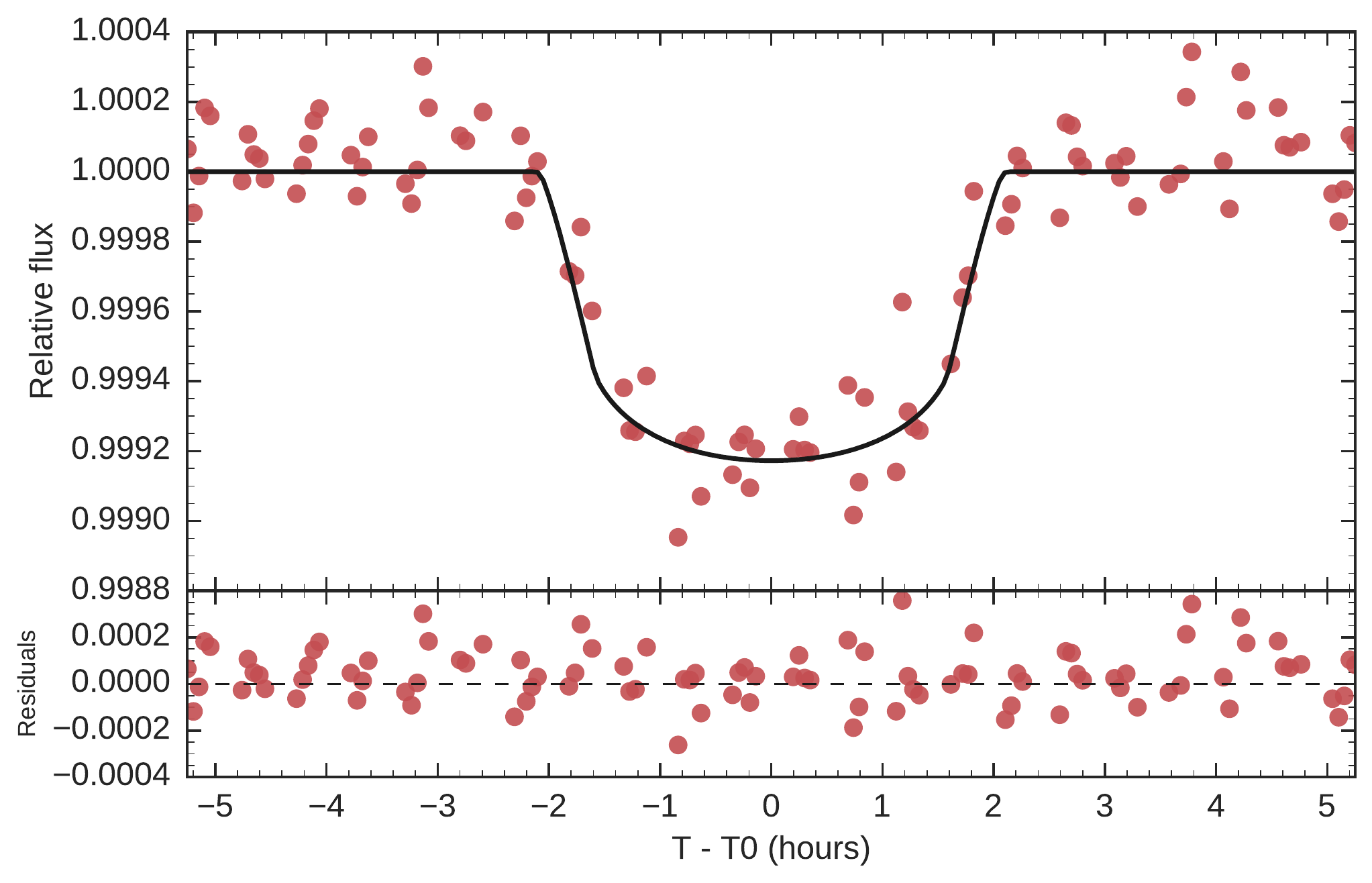}
\caption{Best-fit light curves to planet \object{K2-106\,c}.  The
  light curve has been folded using the orbital period of the planet
  (Table~\ref{tab:values}).}
\label{EPICcTR}
\end{figure}

The phase-folded RV curves of \object{K2-106\,b} and
\object{K2-106\,c} are shown in Figs.~\ref{EPICbRV} and
\ref{EPICcRV}. Figures~\ref{EPICbTR} and \ref{EPICcTR} show
the phase-folded transit light curves. When we use the data obtained by
Sinukoff et al. (\cite{sinukoff17b}), the jitter terms, and the
stellar parameters derived by us, the masses of the two planets are
$\rm M_b=8.36_{-0.94}^{+0.96}$ $\rm M_{\oplus}$, and $\rm
M_c=5.8_{-3.0}^{+3.3}$ $M_{\oplus}$ outer planet, respectively.  The
radius of inner planet is $\rm R_b=1.52\pm0.16$\,$\rm
R_{\oplus}$, and $\rm R_c=2.50_{-0.26}^{+0.27}$\,$\rm R_{\oplus}$ for
the outer planet. With these values, the mean densities are
$13.1_{-3.6}^{+5.4}$ $\rm g\,cm^{-3}$ and $2.0_{-1.1}^{+1.6}$ $\rm
g\,cm^{-3}$ for the two planets, respectively. All the values derived
for the two planets are listed in Table~\ref{tab:values}.  The radii
we have derived are consistent with the values of $\rm
R_p=1.46\pm0.14$\,$\rm R_{\oplus}$ and $\rm R_p=2.53\pm0.14$ \,$\rm
R_{\oplus}$ for the two planets given by Adams et
al. (\cite{adams17}).

With the stellar parameters given in Adams et
  al. (\cite{adams17}), the mass and radius of the inner planet
  becomes $\rm M_b=8.22_{-0.92}^{+0.94}$ $M_{\oplus}$, and $\rm
  R_b=1.45\pm0.15$ $R_{\oplus}$, respectively. With these
  values the density increases to $\rm \rho_b=14.8_{-4.0}^{+6.1}$
  $\mathrm{g\,cm^{-3}}$.  For the outer planet, we find $\rm
  M_c=5.7_{-3.0}^{+3.3}$ $M_{\oplus}$, and $\rm R_c=2.39\pm0.25$\,$\rm
  R_{\oplus}$ , respectively. 

\begin{table}
\caption{K2-106 system parameters.}
\begin{tabular}{l l}
\noalign{\smallskip}
\hline
K2-106 & \\
\hline
$\rm M_*$[$M_{\odot}$]         & $0.945\pm0.063 $ \\ 
$\rm R_*$[$R_{\odot}$]         & $0.869\pm0.088$ \\
$\rm T_{\rm eff}$[$\mathrm{K}$]    & $5470\pm30$ \\
Linear limb-darkening coefficient $u_1$    & $0.41_{-0.12}^{+0.13}$ \\
Quadratic limb-darkening coefficient $u_2$ & $0.25_{-0.12}^{+0.13}$ \\
 $q_1$    & $0.448_{-0.096}^{+0.101}$ \\
 $q_2$    & $0.312_{-0.089}^{+0.091}$ \\
\hline
 & \\
K2-106\,b             & \\                            
\hline
$\rm T_0$ [days]               & $2457394.0114\pm0.0010$ \\ 
Period [days]                  & $0.571292_{-0.000013}^{+0.000012}$ 
\\ 
Impact parameter $b$           & $0.18_{-0.13}^{+0.19}$ \\       
$\rm a/R_*$                    & $2.892_{-0.135}^{+0.089}$ \\       
$\rm R_p/R_*$                  & $0.01601_{-0.00029}^{+0.00031}$ \\       
Radial velocity semi-amplitude $K$ $\rm [ms^{-1}]$ & $6.67\pm0.69$ \\ 
Orbital eccentricity $e$       & 0.0 (fixed) \\
$\rm \sqrt{e} \sin \omega_\star$ &  0.0 (fixed) \\
$\rm \sqrt{e} \cos \omega_\star$ &  0.0 (fixed) \\
Inclination $i$ $\rm [deg]$    & $86.4_{-4.1}^{+2.5}$ \\ 
Orbital semi-major axis $a$ [AU]  & $0.0116\pm0.0013$ \\   
$\rm M_{p}$[$M_{\oplus}$]      & $8.36_{-0.94}^{+0.96}$ \\ 
$\rm R_{p}$[$R_{\oplus}$]      & $1.52\pm0.16$ \\   
$\rm \rho_{p}$[$\mathrm{g\,cm^{-3}}$] & $13.1_{-3.6}^{+5.4}$ \\
$\rm g_{p}$[$\mathrm{cm\,s^{-2}}$]    & $2757_{-396}^{+369}$ \\
$\rm T_{eq}^1$ [K]                   & $2333_{-57}^{+69}$  \\    
$\rm \tau_{14}$ [hours]            & $1.532_{-0.035}^{+0.037}$  \\
$\rm \tau_{12}=\tau_{34}$ [hours]  & $0.0253_{-0.0012}^{+0.0037}$ \\
\hline
 & \\
K2-106\,c          & \\                            
\hline
$\rm T_0$ [days]               & $2457405.73156_{-0.0044}^{+0.0033}$ \\
Period [days]                  & $13.33970_{-0.00096}^{+0.00091}$ \\ 
Impact parameter $b$           & $0.31_{-0.20}^{+0.17}$ \\       
$\rm a/R_*$                    & $26.2_{-2.7}^{+2.4}$ \\       
$\rm R_p/R_*$                  & $0.02632_{-0.00058}^{+0.00075}$ \\       
Radial velocity semi-amplitude $K$ $\rm [ms^{-1}]$ & $1.67_{-0.88}^{+0.99}$ \\
Orbital eccentricity $e$       & $0.18_{-0.12}^{+0.15}$ \\
$\rm \omega$                   & $178_{-74}^{+58}$ \\
$\rm \sqrt{e} \sin \omega_\star$ & $0.01\pm0.25$ \\
$\rm \sqrt{e} \cos \omega_\star$ & $-0.28_{-0.24}^{+0.39}$\\
Inclination $i$ $\rm [deg]$    & $89.35_{-0.46}^{+0.43}$ \\ 
Orbital semi-major axis $a$ [AU] & $0.105_{-0.015}^{+0.015}$ \\   
$\rm M_{p}$[$M_{\oplus}$]       & $5.8_{-3.0}^{+3.3}$ \\ 
$\rm R_{p}$[$R_{\oplus}$]       & $2.50_{-0.26}^{+0.27}$ \\   
$\rm \rho_{p}$[$\mathrm{g\,cm^{-3}}$] & $2.0_{-1.1}^{+1.6}$ \\
$\rm g_{p}$[$\mathrm{cm\,s^{-2}}$]    & $843_{-447} ^{+557}$ \\
$\rm T_{eq}$ [K]                   & $774_{-36}^{+46}$  \\    
$\tau_{14}$ [hours]        & $3.66_{-0.57}^{+0.69}$  \\
\hline
\end{tabular}
\\
The variables are explained in Sect. 3.3.
The jitter terms and the systemic velocities are given in
Tables~\ref{tab:RV1} and~\ref{tab:RV2}. \\
$^1$ Equilibrium temperature $T_{\rm eq}$ derived assuming zero albedo. \\
\label{tab:values}
\end{table}

\subsection{Atmospheric escape rates}
\label{escape}

Because of the relatively similar masses of the
planets and their differing orbital distances, \object{K2-106} is an
excellent laboratory for the study of atmospheric escape.
\object{K2-106\,b} adds to the sample of ultra-short period planets,
such as CoRoT-7b (L\'eger et al. \cite{leger09}) and Kepler-10b (Batalha
et al. \cite{batalha11}), for which the bulk density is suggestive of
an Earth-like composition. Such ultra-short period planets have a
Jeans escape parameter $\Lambda$ below $\approx$20. This is also the
case for \object{K2-106b,} which has $\Lambda=17.1\pm2.6$.

As mentioned in the introduction, the atmosphere of planets with
$\Lambda$ values lower than 20 - 40, depending on the system
parameters, lie in the boil-off regime (Owen \& Wu \cite{owen16};
Cubillos et al.\cite{cubillos17}), where the escape is driven by the
atmospheric thermal energy and low planetary gravity, rather than the
high-energy (XUV) stellar flux. Fossati et al.(\cite{fossati17})
showed that the hydrogen-dominated atmosphere of planets with an
equilibrium temperature higher than 1000\,K, a mass lower than about
5\, $M_{\oplus}$, and a $\Lambda$ value lower than 20-40 should
evaporate completely in less than about 500\,Myr. As indicated by
these theoretical results and by the planet's high bulk density,
\object{K2-106\,b} has probably lost any hydrogen-dominated atmosphere
it may once have had. Because of the very close distance to the host star, the
planet has probably also lost any secondary, likely CO$_2$-dominated,
atmosphere because of the intense stellar radiation (Kulikov et
al. \cite{kulikov06}; Tian \cite{tian09}). The planet could therefore
have been left with a bare rocky surface exposed to the intense
stellar radiation and wind. This may have led to the formation of
surface magma oceans (Leger et al. \cite{leger11}; Miguel et
al. \cite{miguel11}; Demory et al.\cite{demory16}) that outgas and
sputter, in a way  similar from what occurs on Mercury
(Pfleger et al. \cite{pfleger15}). This could create an extended
escaping exosphere composed mostly of heavy refractory elements (Mura
et al. \cite{mura11}).

The parameters of \object{K2-106\,c} are nearly identical to those of
\object{Kepler\,454\,b} (Gettel et al.
\cite{gettel16}). \object{Kepler\,454\,b} is the innermost known
planet of a system that also has a massive planet with an orbital
period of 527\,d. Whether planets like \object{K2-106\,c} and
\object{Kepler\,454\,b} have a rocky core and an extended atmosphere
or if they belong to the elusive class of ``ocean planets''
(L\'eger et al. \cite{leger04}) cannot be deduced from the mass and
radius measurements alone. Further studies are needed to clarify the
situation, but, as mentioned above, it is reasonable to assume that
\object{K2-106\,c} has a rocky core and an extended atmosphere.

The status and evolution of the atmosphere of \object{K2-106\,c} is
also less certain because of the rather large uncertainty in the
planet's mass. We estimated the XUV-driven escape rate based on the
energy-limited formulation of Erkaev et al.(\cite{erkaev07}) and an
XUV (XUV: 1 -- 912\,\AA) flux rescaled from the solar flux (since the
star has a solar-like activity level), obtaining a mass-loss rate
M$_{\rm en}$ of $\rm 2 \times 10^9\,g\,s^{-1}$.  We also employed the
hydrodynamic upper-atmosphere code described by Erkaev et al. (2016),
obtaining a mass-loss rate $\rm M_{hy}$ of $\rm 4\times 10^9\,g\,
s^{-1}$. This and the fact that the planet's $\Lambda$ value is
$25.8\pm9.2$ suggest that the planetary atmosphere may be in the
boil-off regime (Owen \& Wu \cite{owen16}; Fossati et
al. \cite{fossati17}). The parameters relevant to atmospheric escape
are listed in Table~\ref{tab:escape}.

We now assume that the atmosphere of \object{K2-106\,c} is
hydrogen dominated, as suggested by the low bulk density, and that it
is indeed in the boil-off regime. This would imply that the atmosphere
would almost completely escape within a few hundred Myr (Fossati et
al. \cite{fossati17}), which is not compatible with the measured bulk
density and age of the system, which is certainly older than a few
a few hundred Myr.  It is also extremely unlikely that we
have observed the planet during a short-lived transition phase
characterized by an extremely high escape rate. Under the current
assumptions, the most likely possibility is that either the radius
and/or equilibrium temperature are overestimated and/or the mass is
underestimated (Cubillos et al. \cite{cubillos17}). This is the same
situation as considered by Lammer et al.(\cite{lammer16}) for
\object{CoRoT-24\,b} (Alonso et al. \cite{alonso14}) and then extended
by Cubillos et al.(\cite{cubillos17}) to a large sample of low-density
sub-Neptune-mass planets. These authors showed that a radius
overestimation may be caused by the presence of high-altitude
clouds. At the same time, the presence of clouds would also imply that
the equilibrium temperature may have been overestimated because this
would increase the albedo (see Cubillos et al. \cite{cubillos17} for
more details). A better understanding of the loss processes would be
possible with a higher accuracy in the mass and radius determinations
of the planet and the star.

\begin{table}
\caption{Atmospheric escape parameters}
\begin{tabular}{l r}
\hline
\noalign{\smallskip}
K2-106\,b             & \\                            
\hline
Restricted Jeans escape parameter $\Lambda$      &    $17.1\pm2.6$ \\
Roche-lobe radius [$\rm R_{\oplus}$]  &     4 \\
$\rm F_{XUV}$ [$\rm erg\,cm^{-2}\,s^{-1}$]  & 11500 \\
Escape rate [$\rm s^{-1}$]            & $2.1\times 10^{+33}$ \\
$\rm F_p/F_{\oplus}^1$                & 3500 \\ 
\hline
 & \\
K2-106\,c             & \\                            
\hline
Restricted Jeans escape parameter $\Lambda$      & $25.8\pm9.2$  \\
Roche-lobe radius [$\rm R_{\oplus}$]  & 33 \\
$\rm F_{XUV}$ [$\rm erg\,cm^{-2}\,s^{-1}$]  & 154 \\
Escape rate   [$\rm s^{-1}$]          & $6.6\times 10^{+31}$ \\
$\rm F_p/F_{\oplus}^1$                & 52 \\
\hline 
\end{tabular}
\label{tab:escape}
\\
$^1$ Ratio of the stellar flux received by the planet compared to
Earth.
\end{table}

\section{Discussion and conclusions}
\label{sectIV}

We have determined the masses of the planets \object{K2-106\,b} and
\object{K2-106\,c}.  \object{K2-106\,b} is a low-mass
ultra-short-period planet.  Table~\ref{tab:ultrashort} gives an
overview of the known planets of this type. Other planets of this
class are \object{CoRoT-7b} (L\'eger et al. \cite{leger09}),
\object{55\,Cnc\,e} (Winn et al. \cite{winn11}), Kepler-10b (Batalha
et al. \cite{batalha11}), Kepler-78b (Sanchis-Ojeda et
al. \cite{sanchis13}), WASP-47e (Dai et al. \cite{dai15}; Sinukoff et
al. (\cite{sinukoff17a}), and perhaps also the planet candidates of
the sdB KIC 05807616 (Charpinet et al. \cite{charpinet11}).  Because
\object{KOI 1843.03} has an orbital period of only 4.425 hr, Rappaport
et al.  (\cite{rappaport13}) concluded that this planet must have a
density higher than $\rm \rho=7\,g\,cm^{-3}$. Adams et
al. (\cite{adams16}) recently published a list of 19 additional
planet candidates with orbital periods shorter than one day.  One of
these is \object{EPIC\,203533312,} which has an orbital period of 4.22
hr. If confirmed as a planet, its density must be higher than $\rm
\rho=8.9\,g\,cm^{-3}$. 

\begin{figure}
\includegraphics[height=.26\textheight,angle=0.0]{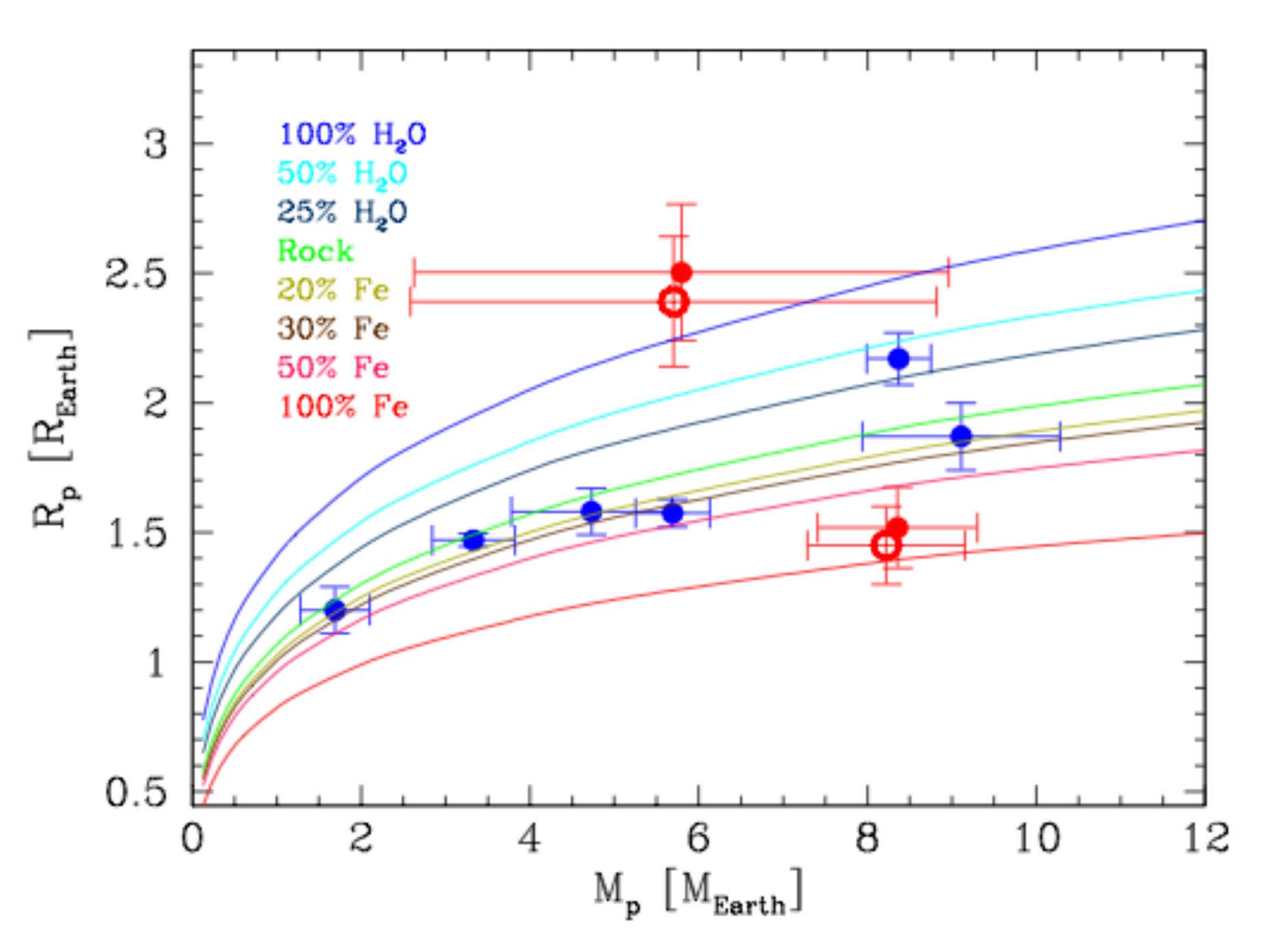}
\includegraphics[height=.25\textheight,angle=0.0]{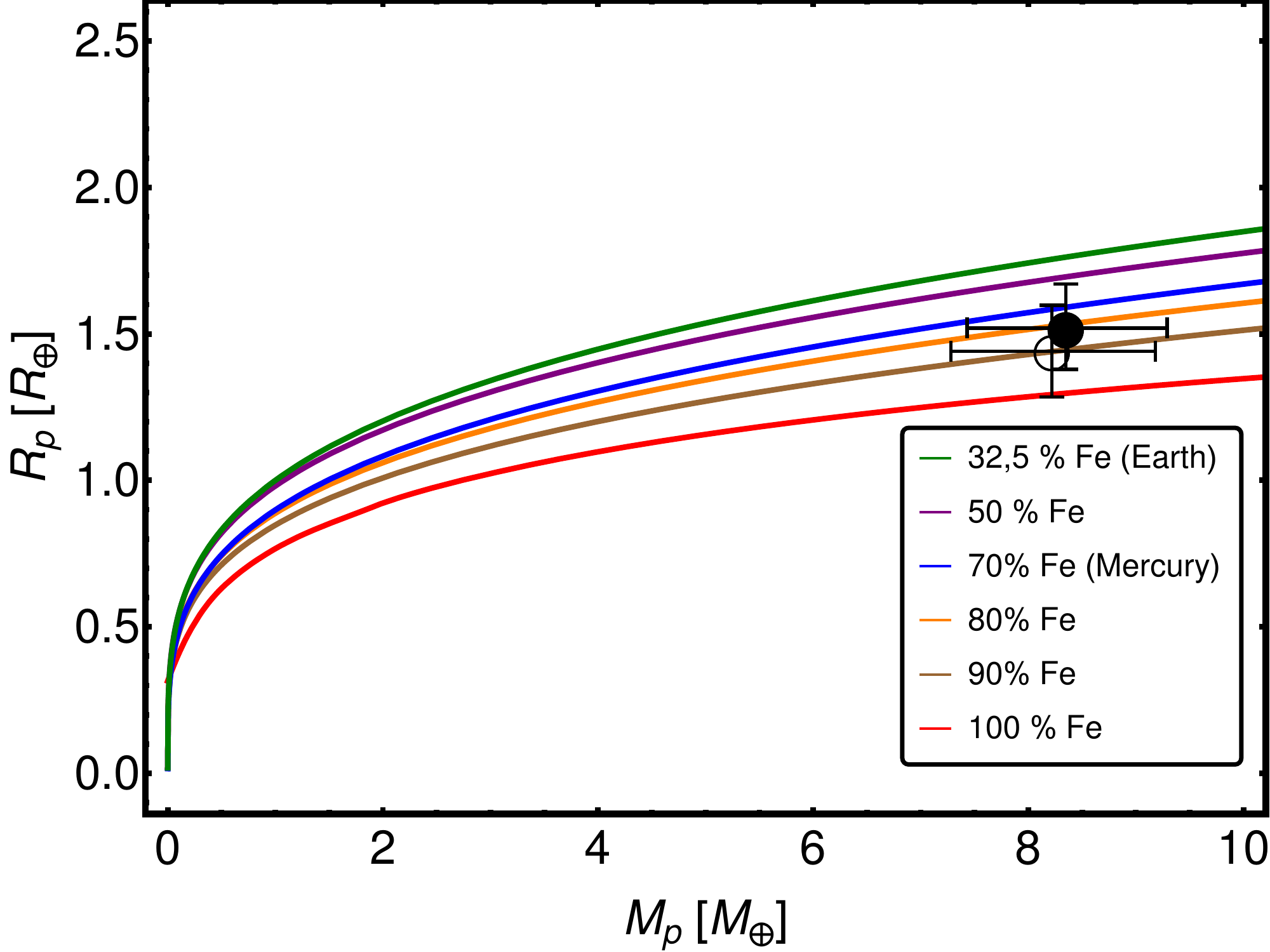}
\caption{Upper panel: mass-radius relation for low-mass ultra-short planets.   \object{K2-106\,b,c} are the red filled symbols obtained using
  the stellar parameters derived by us. The open red symbol shows the
  same, but for the parameters given by Adams et al. (\cite{adams17}).
  The blue symbols are all other known ultra-short period planets.
  Lower panel: detailed position of \object{K2-106\,b,c} using models calculated by
  us. The iron core contains about 80\% of the mass of the planet.}
\label{EPICcom}
\end{figure} 

\begin{table*}
\caption{Low-mass ultra-short-period planets with known densities}
\begin{tabular}{l c c c c c l }
\hline
\noalign{\smallskip}
planet & orbital    & mass        & radius      & density & multiple     & ref. \\
       & period [d] & [$M_{\oplus}$] & [$R_{\oplus}$] & [$\rm g\,cm^{-3}$] & system$^1$ & \\
\hline
KOI 1843.03  & 0.18 & $\geq 0.46^3$ & $0.45_{-0.05}^{+0.08}$ & $\geq 7^3$  & NO & Rappaport et al.  \cite{rappaport13} \\ 
Kepler-78\,b       & 0.36 & $1.69\pm0.41$ & $1.20\pm0.09$          & $5.3\pm1.8$          & NO & Howard et al. \cite{howard13} \\
K2-106\,b          & 0.57 & $8.36_{-0.94}^{+0.96}$ & $1.52\pm0.16$ & $13.1_{-3.6}^{+5.4}$ & YES & this work \\
55 Cnc\,e$^2$      & 0.74 & $8.63\pm0.35$ & $2.00\pm0.14$          & $5.9^{+1.5}_{-1.1}$  & YES & Winn et al. \cite{winn11}\\
55 Cnc\,e$^2$      & 0.74 & $8.37\pm0.38$ & $2.17\pm0.10$          & $4.5\pm0.20$  & YES & Winn et al. \cite{endl12}\\    
WASP-47\,e$^2$     & 0.79 & $12.2\pm3.7$  & $1.817\pm0.065$        & $11.3\pm3.6$         & YES & Dai et al. \cite{dai15}\\
WASP-47\,e$^2$     & 0.79 & $9.11\pm1.17$ & $1.87\pm0.13$          & $7.63\pm1.90$        & YES & Sinukoff et al. \cite{sinukoff17a}\\
Kepler-10\,b       & 0.84 & $3.33\pm0.49$ & $1.47^{+0.03}_{-0.02}$ & $5.8\pm0.8$          & YES & Dumusque et al. \cite{dumusque14} \\
CoRoT-7\,b         & 0.85 & $4.73\pm0.95$ & $1.58\pm0.09$          & $6.61\pm1.72$        & YES & Haywood et al. \cite{haywood14}  \\
HD3167\,b    & 0.95 & $5.69\pm0.44$ & $1.574\pm0.054$          & $8.00^{+1.10}_{-0.98}$ & YES & Gandolfi et al. \cite{gandolfi17} \\
\hline
\end{tabular}
\label{tab:ultrashort}
\\
$^1$ ``NO'' means that no other planet is known.  \\
$^2$ Two measurements were obtained for this planet. \\
$^3$ Lower limit of mass and density estimated from orbital period and radius. \\
\end{table*}

The upper panel in Fig.~\ref{EPICcom} shows the mass-radius
  relation for low-mass ultra-short planets together with various
  compositions taken from Zeng et al. (\cite{zeng16}). The filled red
  symbols are the values for \object{K2-106\,b} and \object{K2-106\,c}
  using our stellar parameters, the open symbol those for
  the stellar parameters given by Adams et al. (\cite{adams17}). In
  both cases, \object{K2-106\,b} is located between the lines with 50\%
  and 100\% iron composition.  How robust is this conclusion that
  \object{K2-106\,b} is metal rich? When we use models published by
  Fortney et al. (\cite{fortney07}) or Wurm et al. (\cite{wurm13}),
  we obtain the same results. Thus, regardless of which set of stellar
  parameters we use, whether we include the jitter term, and
regardless of which models we use, in all cases we reach the conclusion that the
  iron core contains more than half of the mass of the planet. The
  conclusion that this planet is metal rich is therefore robust. 
In order to constrain the composition more precisely, we used our own 
planetary models. These are two-layer models of iron and silicates 
($\rm MgSiO_3$) based on the model of Wagner (\cite{wagner11}). 
Using these models, we find that an iron core
  contains $80_{-30}^{+20}$\% of the mass of the planet. The
  composition is Mercury-like rather than Earth-like. This is very
  interesting because other ultra-short planets seem to have an
  Earth-like composition.  The high metal content of the planet is particularly 
surprising because the host star has solar metallicity
 (Table~\ref{tab:star}).  The unusual composition of
  \object{K2-106\,b also} shows that rocky planets are more diverse
  than previously thought, and it can provide important clues of how such 
metal-rich planets form.

As pointed out by Alessi et al. (\cite{alessi17}), the variety of
chemical compositions observed for giant planets could be caused by
variations in metallicities of their host stars or by the accretion of
material at different locations in disks around stars with similar
compositions.  Thorngren et al. (\cite{thorngren16}) have studied the
relation between the planetary heavy-element mass and the total planet
mass for planets in the mass range between 20 and 3000 $\rm
M_{\oplus}$ (0.07-10 $\rm M_{Jup}$). They found a clear correlation
between the two, in the sense that heavy-element mass increases with
the mass of the host star.

Measuring abundances for low-mass ultra-short period planets is
particularly interesting because the densities of planets without
atmospheres constrain the formation of planets (Raymond et
al. \cite{raymond13}).  For example, close-in planets with a high
water content are likely to have formed at a larger distance from the host
star and then migrated inward. An interesting result emerging from the
models calculated by Lopez (\cite{lopez16}) is that planets
that receive around 2800 times the stellar flux of Earth can keep
substantial water envelopes. Such planets would have $R_p\geq
2\,R_{\oplus}$.  This is not the case for \object{K2-106\,b}, which
means that it must have formed from water-poor material inside the
snow line.

When we apply the same logical argument to the iron versus silicate
contents, it would mean that \object{K2-106\,b} formed from metal-rich
material. In the solar system, Mercury is 70\% metal and 30\%
silicate, which implies a similar formation scenario as for
\object{K2-106\,b}. Therefore it is reasonable to consider similar
formation scenarios for \object{K2-106\,b} and Mercury.  In this
respect, it is interesting to note that Wurm et al. (\cite{wurm13})
argued that the high iron abundance of Mercury is due to the
photophoresis in the protoplanetary disk and not the result of a
giant impact, as was previously thought. Photophoresis is a process in
which iron and silicates are separated in the disk. Iron ends up in
the very innermost part of the disk, with silicates at somewhat larger
distances.  At the current location, the temperature is also higher
than the silicate evaporation temperature in the disk, which is in the
range between 1300 and 1450 K (Gail \cite{gail98}).  Is it possible
that ultra-short period planets form close to the star from iron-rich
material?  A planet of 8 $\rm M_{\oplus}$, forming at 0.012 AU,
certainly requires much material in the disk, but according to the
model published by Hasegawa \& Pudritz (\cite{hasegawa13}), it is
possible to form such planets close to the star.  If ultra-short
period planets are forming close to the star, many of them should be
iron-rich. Clearly, more research in this field is needed, but the
results so far obtained show that studies of ultra-short period
planets can give us key information on how and where low-mass planets
form. 

Another interesting aspect of the \object{K2-106\,b,c} system is that
the masses of the two planets are relatively similar, but the densities
are very different.  Since the mass and radius measurements for both
planets are affected in the same way by the systematic uncertainties
of the stellar parameters, the diversity of exoplanets (e.g., Hatzes \&
Rauer \cite{hatzes15}) cannot be entirely explained by problems in the
determination of the stellar parameters. As shown in
Fig.~\ref{EPICcom}, the density of \object{K2-106\,c} is consistent
with a planet composed of 50\% rock and 50\% ice. However, as pointed
out above, other planets like this one consist of a rocky core with a
hydrogen atmosphere (Chen et al. \cite{chen17}). Since the mass is
$\rm M_c=5.8_{-3.0}^{+3.3}$ $M_{\oplus}$ and the ratio of the stellar
flux received by the planet compared to Earth is $\rm
F_p/F_{\oplus}\sim 52,$ there is no reason why it could not have a
hydrogen atmosphere since at about 800 K a 50\% ice content is not very
plausible.  The difference in atmospheric loss rates, which we have
discussed in Sect. \ref{escape}, explains why the inner planet has
no hydrogen atmosphere, while the other planet is likely to have
one. However, as we pointed out above, the fact that $\Lambda$ is in
the interesting regime between 20 and 40 makes \object{K2-106\,c} an
ideal target for future studies of atmospheric escape,
particularly because the host star is much brighter than that of
\object{CoRoT-24\,b} (Alonso et al. \cite{alonso14}).

We conclude that \object{K2-106}
(\object{EPIC\,220674823},\object{TYC\,608-458-1}) is an interesting
system that deserves further study. The accuracy with which the radius
of the star and thus also the planets can be determined will increase
once Gaia collects more data. A logical step for future work is to
search for the extended escaping exosphere atmosphere of
\object{K2-106\,b} that has been suggested by Mura et
al. (\cite{mura11}) by obtaining spectroscopic transit observations
in a similar way as for CoRoT-7b (Guenther et al.
\cite{guenther11}).

\begin{acknowledgements}

This work was generously supported by the Th\"uringer Ministerium
f\"ur Wirtschaft, Wissenschaft und Digitale Gesellschaft and the
Deutsche Forschungsgemeinschaft (DFG) under the project GU 464/20-1.
HD acknowledges support by grant ESP2015-65712-C5-4-R of the Spanish
Secretary of State for R\&D\&i (MINECO).  LF and DK acknowledge the
Austrian Forschungsf\"orderungsgesellschaft FFG project
``TAPAS4CHEOPS'' P853993.  MF and CMP acknowledge generous support
from the Swedish National Space Board.  This work was supported by
Japan Society for Promotion of Science (JSPS) KAKENHI Grant Number
JP16K17660, and by the Astrobiology Center Project of National
Institutes of Natural Sciences (NINS) (Grant Number JY280092).
Results based in part on data collected at Subaru Telescope, which is
operated by the National Astronomical Observatory of Japan.  This
paper includes data gathered with the 6.5 meter Magellan Telescopes
located at Las Campanas Observatory, Chile.  Partly based on
observations made with ESO Telescopes at the La Silla Paranal
Observatory under programme ID 098.C-0860(A).  Australian access to
the Magellan Telescopes was supported through the National
Collaborative Research Infrastructure Strategy of the Australian
Federal Government.  Also based in part on observations made with the
Italian Telescopio Nazionale Galileo (TNG) operated on the island of
La Palma by the Fundaci\'on Galileo Galilei of the INAF (Istituto
Nazionale di Astrofisica) at the Spanish Observatorio del Roque de los
Muchachos of the Instituto de Astrof\'\i sica de Canarias (IAC).  Also
partly based on observations made with the Nordic Optical Telescope,
operated by the Nordic Optical Telescope Scientific Association at the
Observatorio del Roque de los Muchachos, La Palma, Spain, of the
Instituto de Astrof\'\i sica de Canarias.  This work has made use of
data from the European Space Agency (ESA) mission Gaia
({https://www.cosmos.esa.int/gaia}), processed by the Gaia Data
Processing and Analysis Consortium (DPAC,
{https://www.cosmos.esa.int/web/gaia/dpac/consortium}). Funding for
the DPAC has been provided by national institutions, in particular the
institutions participating in the {\it Gaia} Multilateral Agreement.
This research has made use of the SIMBAD database, operated at CDS,
Strasbourg, France. We are grateful to Jorge Melendez, Fran\c{c}ois
Bouchy, and Xavier Bonfils, who kindly agreed to exchange HARPS time
with us. We thank the NOT staff members for their valuable support
during the observations. DG gratefully acknowledges the financial
support of the \emph{Programma Giovani Ricercatori -- Rita Levi
  Montalcini -- Rientro dei Cervelli (2012)} awarded by the Italian
Ministry of Education, Universities and Research (MIUR).

\end{acknowledgements}


\begin{thebibliography}{}

\bibitem[2017]{adams17} Adams, E.R., et al. \ 2017, AJ 153, 82

\bibitem[2016]{adams16} Adams, E.R., Jackson, B., \& Endl, M.\ 2016,
  \aj, 152, 47

\bibitem[2017]{alessi17} Alessi, M., Pudritz, R.~E., \& Cridland,
  A.~J.\ 2017, \mnras, 464, 428 Alessi

\bibitem[2014]{alonso14} Alonso, R., Moutou, C., Endl, M., et
  al.\ 2014, \aap, 567, A112

\bibitem[1996]{baranne96} Baranne, A., et al.\ 1996, \aaps, 119, 373

\bibitem[2016]{barragan16} Barrag{\'a}n, O., Grziwa, S., Gandolfi, D.,
  et al.\ 2016, \aj, 152, 193

\bibitem[2017]{barragan17} Barrag{\'a}n, O., Gandolfi, D., \& 
Antoniciello, G.\ 2017, Astrophysics Source Code
Library, ascl:1707.003

\bibitem[2011]{batalha11} Batalha, N.M., Borucki, W.J., Bryson, S.T.,
  et al.\ 2011, \apj, 729, 27

\bibitem[1996]{butler96} Butler, R.~P., Marcy, G.~W., Williams, E.,
  et al.\ 1996, \pasp, 108, 500

\bibitem[2015]{berta15} Berta-Thompson, Z.K., Irwin, J., Charbonneau,
  D., et al.\ 2015, \nat, 527, 204

\bibitem[2016]{bourrier16} Bourrier, V., Lecavelier des Etangs, A.,
  Ehrenreich, D., Tanaka, Y.~A., \& Vidotto, A.~A.\ 2016, \aap, 591,
  A121

\bibitem[2010]{bruntt10} Bruntt, H., Bedding, T.~R., Quirion, P.-O.,
  et al.\ 2010, \mnras, 405, 1907

\bibitem[2012]{carter12} Carter, J.A., Agol, E., Chaplin, W.J., et
  al.\ 2012, Science, 337, 556

\bibitem[2017]{chen17} Chen, G., Guenther, E.~W., Palle, E., et
  al.\ 2017, arXiv:1703.01817

\bibitem[2011]{charpinet11} Charpinet, S., Fontaine, G., Brassard, P.,
  et al.\ 2011, \nat, 480, 496

\bibitem[2011]{claret11} Claret, A., \& Bloemen, S.\ 2011, \aap, 529,
  A75

\bibitem[2012]{cosentino12} Cosentino, R., Lovis, C., Pepe, F., et
  al.\ 2012, \procspie, 8446, 84461V

\bibitem[2006]{crane06} Crane J. D., Shectman S. A., Butler R. P. et
  al 2010 Proc. SPIE 7735 773553

\bibitem[2008]{crane08} Crane J. D., Shectman S. A., Butler R. P.,
  Thompson I. B. and Burley G. S. 2008

\bibitem[2010]{crane10} Crane J. D., Shectman S. A. and Butler
  R. P. 2006 Proc. SPIE 6269 626931

\bibitem[2017]{cubillos17} Cubillos, P., Erkaev, N.~V., Juvan, I., et
  al.\ 2017, \mnras, 466, 1868

\bibitem[2015]{dressing15} Dressing, C.D., Charbonneau, D., Dumusque,
  X., et al.\ 2015, \apj, 800, 135

\bibitem[2006] {dadilva06} da Silva, L., Girardi, L., Pasquini, L., et
  al.\ 2006, \aap, 458, 609

\bibitem[2015]{dai15} Dai, F., Winn, J.~N., Arriagada, P., et
  al.\ 2015, \apjl, 813, L9

\bibitem[2016]{demory16} Demory, B.-O., Gillon, M., de Wit, J., et
  al.\ 2016, Nature, 532, 207

\bibitem[2014]{dumusque14} Dumusque, X., Bonomo, A.~S., Haywood,
  R.~D., et al.\ 2014, \apj, 789, 154

\bibitem[2015]{ehrenreich15} Ehrenreich, D., Bourrier, V., Wheatley,
  P.~J., et al.\ 2015, \nat, 522, 459

\bibitem[2017]{eigmueller17} Eigm{\"u}ller, P., Gandolfi, D., Persson,
  C.~M., et al.\ 2017, \aj, 153, 130

\bibitem[2012]{endl12}Endl, M., Robertson, P., Cochran, W.~D., et
  al.\ 2012, \apj, 759, 19

\bibitem[2007]{erkaev07} 
  Erkaev, N.~V., Kulikov, Y.~N., Lammer, H., et al.\ 2007, \aap, 472, 329 

\bibitem[2007]{fortney07} Fortney, J.~J., Marley, M.~S., \& Barnes,
  J.~W.\ 2007, \apj, 659, 1661

\bibitem[2017]{fossati17} Fossati, L., Erkaev, N.~V., Lammer, H., et
  al.\ 2017, \aap, 598, A90

\bibitem[1999]{frandsen99} Frandsen, S. \& Lindberg, B. 1999, in
  ``Astrophysics with the NOT'', proceedings Eds: Karttunen, H. \&
  Piirola, V., anot.  conf, 71

\bibitem[2016a]{gaia2016a}Gaia Collaboration et al. (2016a) Gaia Data
  Release 1. Summary of the astrometric, photometric, and survey
  properties. \aap 595, A2.

\bibitem[2016b]{gaia2016b} Gaia Collaboration et al. (2016b) The Gaia
  mission. \aap 595, A1.

\bibitem[1998]{gail98} Gail, H.-P.\ 1998, \aap, 332, 1099

\bibitem[2008]{gandolfi08} Gandolfi, D., Alcal{\'a}, J.~M., Leccia,
  S., et al.\ 2008, \apj, 687, 1303-1322

\bibitem[2015]{gandolfi15} Gandolfi, D., Parvianinen, H., Deeg, H. J.,
  et al. 2015, \aap, 576, A11

\bibitem[2017]{gandolfi17} Gandolfi, D., Barrag{\'a}n, O., 
    Hatzes, A.P., et al.\ 2017, \aj, 154, 123 

\bibitem[2016]{gettel16} Gettel, S., Charbonneau, D., Dressing, C.~D.,
  et al.\ 2016, \apj, 816, 95

\bibitem[2000]{girardi00} Girardi, L., Bertelli, G., Bressan, A., et
  al.\ 2002, \aap, 391, 195

\bibitem[2002]{girardi02} Girardi, L., Bressan, A., Bertelli, G., \&
  Chiosi, C.\ 2000, \aaps, 141, 371

\bibitem[2011]{guenther11} Guenther, E.~W., Cabrera, J., Erikson, A.,
  et al.\ 2011, \aap, 525, A24

\bibitem[2014]{haywood14} Haywood, R.~D., Collier Cameron, A., Queloz,
  D., et al.\ 2014, \mnras, 443, 2517

\bibitem[2015]{hatzes15} Hatzes, A.~P., \& Rauer, H.\ 2015, \apjl,
  810, L25

\bibitem[2013]{hasegawa13} Hasegawa, Y. \& Pudritz, R. E. 2013, \apj,
  778, 78

\bibitem[2013]{howard13} Howard, A.~W., Sanchis-Ojeda, R., Marcy,
  G.~W., et al.\ 2013, \nat, 503, 381

\bibitem[2016]{huber16} Huber, D., Bryson, S. T., Haas, M. R., et
  al. 2016, \apjs, 224, 2

\bibitem[2016]{johnson16} Johnson, M.~C., Gandolfi, D., Fridlund, M.,
  et al.\ 2016, \aj, 151, 171

\bibitem[2010]{kipping10} Kipping, D.\,M. 2010, \mnras, 408, 1758

\bibitem[2013]{kipping13} Kipping, D.~M.\ 2013, \mnras, 435, 2152 

\bibitem[2006]{kulikov06} Kulikov, Y.~N., Lammer, H., Lichtenegger,
  H.~I.~M., et al.\ 2006, P\&SS, 54, 1425

\bibitem[2013]{kurucz13} Kurucz, R. L. 2013, ATLAS12: Opacity sampling
  model atmosphere program, Astrophysics Source Code Library

\bibitem[2014]{lammer14} Lammer, H., et al.\ 2014, \mnras, 439, 3225;

\bibitem[2016]{lammer16} Lammer, H., Erkaev, N.~V., Fossati, L., et
  al.\ 2016, \mnras, 461, L62

\bibitem[2016]{lanza16} Lanza, A.~F., Molaro, P., Monaco, L., \&
  Haywood, R.~D.\ 2016, \aap, 587, A103

\bibitem[2004]{leger04} L{\'e}ger, A., Selsis, F., Sotin, C., et
  al.\ 2004, \icarus, 169, 499

\bibitem[2009]{leger09} L\'eger, A., Rouan, D., Schneider, J., Barge,
  P., Fridlund, M., Samuel, B., Ollivier, M., Guenther, E., Deleuil,
  M., Deeg, H. J. et al.\ 2009, \aap, 506, 287

\bibitem[2011]{leger11} L{\'e}ger, A., Grasset, O., Fegley, B., et
  al.\ 2011, \icarus, 213, 1

\bibitem[2016b]{lindegren16} Lindegren et al. \ 2017 , \aap, 595,A4

\bibitem[2016]{lopez16} Lopez, E.~D.\ 2016, arXiv:1610.01170 

\bibitem[2016]{lundkvist16} Lundkvist, M.~S., Kjeldsen, H., Albrecht,
  S., et al.\ 2016, Nature Communications, 7, 11201

\bibitem[2002]{mandel02} Mandel, K., \& Agol, E.\ 2002, \apjl, 580, L171

\bibitem[2003]{mayor03} Mayor, M., Pepe, F., Queloz, D., et al. 2003,
  Msngr, 114, 20

\bibitem[2010a]{meunier10a} Meunier, N., Lagrange, A.-M., \& Desort,
  M.\ 2010a, \aap, 519, A66

\bibitem[2010b]{meunier10b} Meunier, N., Desort, M., \& Lagrange,
  A.-M.\ 2010b, \aap, 512, A39

\bibitem[2011]{miguel11} Miguel, Y., Kaltenegger, L., Fegley, B., \&
  Schaefer, L.\ 2011, \apjl, 742, L19

\bibitem[2015]{motalebi15} Motalebi, F., Udry, S., Gillon, M., et
  al.\ 2015, \aap, 584, A72

\bibitem[2011]{mura11} Mura, A., Wurz, P., Schneider, J., et
  al.\ 2011, \icarus, 211, 1

\bibitem[2017]{narita17} Narita, N., Hirano, T., Fukui, A., et
  al.\ 2017, \pasj,

\bibitem[2002]{noguchi02} Noguchi, K., Aoki, W., Kawanomoto, S., et
  al.\ 2002, \pasj, 54, 855)

\bibitem[2016]{owen16} Owen, J.~E., \& Wu, Y.\ 2016, \apj, 817, 107

\bibitem[2002]{pepe02} Pepe, F., Mayor, M., Galland, F., et al. 2002,
  \aap, 388, 632

\bibitem[2015]{pfleger15} Pfleger, M., Lichtenegger, H.~I.~M., Wurz,
  P., et al.\ 2015, P\&SS, 115, 90

\bibitem[2013]{raymond13} Raymond, S.~N., Barnes, R., \& Mandell,
  A.~M.\ 2008, \mnras, 384, 663

\bibitem[2013]{rappaport13}Rappaport, S., Sanchis-Ojeda, R., Rogers,
  L.~A., Levine, A., \& Winn, J.~N.\ 2013, \apjl, 773, L15

\bibitem[2006]{saar06}Saar, S. H. 2006, in BAAS 38, AAS/Solar Physics
  Division Meeting 37, 240

\bibitem[2013]{sanchis13} Sanchis-Ojeda, R., Rappaport, S., Winn,
  J.~N., et al.\ 2013, \apj, 774, 54

\bibitem[2014] {sanchis14} Sanchis-Ojeda, R., Rappaport, S., Winn,
  J.~N., et al.\ 2014, \apj, 787, 47

\bibitem[2002]{sato02} Sato, B., et al. 2002, Development of Iodine
  Cells for the Subaru HDS and the Okayama HIDES: II. New Software for
  Precise Radial Velocity Measurements, PASJ, 54, 873

\bibitem[2017a]{sinukoff17a} Sinukoff, E., Howard, A.W., Petigura,
  E.~A., et al.\ 2017a, \aj, 153, 70

\bibitem[2017b]{sinukoff17b} Sinukoff, E., Howard, A.W., Petigura,
  E.A., et al.\ 2017b, arXiv:1705.03491

\bibitem[2012]{tajitsu12} Tajitsu, A., Aoki, W., \& Yamamuro, T. 2012,
  PASJ, 64, 77

\bibitem[2009]{tian09} Tian, F.\ 2009, \apj, 703, 905

\bibitem[2014]{telting14} Telting, J. H., Avila, G., Buchhave, L., et
  al. 2014, Astronomische Nachrichten, 335, 41

\bibitem[2016]{thorngren16} Thorngren, D.~P., Fortney, J.~J.,
  Murray-Clay, R.~A., \& Lopez, E.~D.\ 2016, \apj, 831, 64

\bibitem[2014]{vanderburg14} Vanderburg, A., \& Johnson,
  J.~A.\ 2014, \pasp, 126, 948

\bibitem[2016]{vanderburg16} Vanderburg, A., Latham, D.~W.,
  Buchhave, L.~A., et al.\ 2016, \apjs, 222, 14

\bibitem[1996]{valenti96} Valenti, J.A. \& Piskunov, N. 1996, \aaps,
  118, 595

\bibitem[2005]{valenti05} Valenti, J.A. \& Fischer, D.A. 2005, \apjs,
  159, 141

\bibitem[2011]{wagner11} Wagner, F.W., Sohl, F., Hussmann, H., Grott,
  M., \& Rauer, H.\ 2011, \icarus, 214, 366

\bibitem[2011]{winn11} Winn, J.~N., Matthews, J.~M., Dawson, R.~I., et
  al.\ 2011, \apjl, 737, L18

\bibitem[2013]{wurm13} Wurm, G., Trieloff, M., \& Rauer, H.\ 2013, \apj, 769, 78 

\bibitem[2016]{zeng16} Zeng, L., Sasselov, D.~D., \& Jacobsen,
  S.~B.\ 2016, \apj, 819, 127

\end{thebibliography}
\end{document}